\pdfoutput=1
\documentclass[11pt]{article}
\usepackage{textcomp}
\usepackage{chet}
\usepackage{graphicx,epsfig,color}

\title{Gravity duals for defect quivers in the Veneziano limit}
\author{
Niko Jokela$^{\dag,}$\emailV{niko.jokela@helsinki.fi},
Jos\'e Manuel Pen\'in,$^{\dag,}$\emailV{jose.peninascariz@helsinki.fi} and
Konstantinos Christos S.~Rigatos$^{\ddag, *,}$\emailV{konstantinoschristos.rigatos@colorado.edu}
}
\affiliation{
$^{\dag}$Department of Physics, and Helsinki Institute of Physics, P.O.Box 64 \\ FIN-00014 University of Helsinki, Finland \\ 
$^{\ddag}$Department of Physics and Center for Theory of Quantum Matter 390 UCB \\
University of Colorado Boulder, CO 80309, USA\\
$^{*}$School of Physics \& Astronomy and STAG Research Centre \\ 
University of Southampton, Highfield, SO17 1BJ, United Kingdom
}
\abstract{ 
We construct gravity duals to supersymmetric gauge theories in the presence of unquenched flavor hypermultiplets in the fundamental representation of the gauge group living on the (1+1)-dimensional defect. This configuration is given by the intersection of two sets of D3-branes. Working in the Veneziano limit with large number of colors and flavors we are able to find a closed set of equations describing the dual geometry. We briefly discuss the corresponding solutions for massless flavors as well as in the small flavor limit of massive hypermultiplets. Interestingly, the 1/4-BPS supergravity solutions are generically only asymptotically anti de-Sitter, the dilaton does not vary with the holographic radial coordinate. This implies that the classical Type IIB supergravity solutions remain trustworthy descriptions for the gauge theories from the deep IR to the far UV.   
}
\preprint{HIP-2021-52/TH}

\begin{document}
\maketitle
\toc
\newpage

\newsec{Introduction}

The gauge/string duality \cite{Maldacena:1997re, Witten:1998qj, Gubser:1998bc} is a bona fide paradigm of the holographic principle. It posits that certain field theories are equivalently described by string theory in asymptotically anti de-Sitter (AdS) spacetimes. The prototype proposal related type IIB string theory in $AdS_5 \times S^5$ to the four-dimensional $\mathcal{N}=4$ super Yang-Mills (SYM) and has been generalized to more general gauge/gravity dual pairs in various dimensions, by considering different D$p$-branes \cite{Itzhaki:1998dd}. The essence of the duality is distilled in the twofold nature of D-branes; hypersurfaces on which open strings can have their endpoints, as well as solitonic solutions of supergravity.

Amongst the many visages and developments of the duality, two of them have been of great significance in order to bring holography much closer to the realistic gauge theories that appear in nature. The first has to do with the establishment of holographic examples containing fields that transform in the fundamental representation of the gauge group. An early attempt describing a field theory with fundamental matter holographically can be found in \cite{Fayyazuddin:1998fb, Aharony:1998xz, Bertolini:2001qa} by considering the conformal ${\mathcal{N}}=2$ Sp($2N_c$) theory in terms of D3-branes near F-theory singularities; $AdS_5 \times S^2/Z_2$ string theory. The second development had to do with constructing less supersymmetric exemplars, see for instance \cite{Grana:2005sn}.

The inclusion of flavor degrees of freedom can be obtained by considering additional (flavor) D-branes in the bulk, such that strings can have one endpoint on the D-branes carrying the color degrees of freedom and another on the new D-branes that carry flavors \cite{Karch:2002sh, Karch:2000gx}. The aforesaid open strings are holographically dual to fields that transform in the fundamental representation of the gauge group, and we call these fields quarks. In a first approximation one can consider the number of flavors to be small, which amounts to adding the flavor branes as probes in the geometry. In the field theory language, this is the so-called the quenched limit. In that limit, we can safely ignore the dynamical effects due to the presence of quarks, or equivalently the backreaction of the flavor branes to the near-horizon supergravity solution. If we consider a theory with a number of color degrees of freedom, $N_c$, and a number, $N_f$, of flavors then the quenched limit corresponds to $\tfrac{N_f}{N_c} \rightarrow 0$. There are many interesting results in this limit, such as meson spectroscopy \cite{Kruczenski:2003be, Myers:2006qr, Arean:2006pk, Abt:2019tas, Nakas:2020hyo}, the description of chiral symmetry breaking \cite{Kruczenski:2003uq, Babington:2003vm} or field theories furnishing spatial defects of different dimensionalities \cite{DeWolfe:2001pq, Constable:2002xt}. We refer the interested reader to \cite{Erdmenger:2007cm,Ramallo:2013bua} for more discussion and a detailed account of some of these developments. 

While the probe limit has offered much insight, it is desirable to be able to go beyond that. There are both theoretical and phenomenological reasons to be after this program. The quantum effects of the quark fields have consequences on the existence of Seiberg dualities \cite{Seiberg:1994pq} and conformal points (or conformal windows) in the phase space of the theory \cite{Banks:1981nn}. With an eye towards phenomenology, the color charge screening is, also, due to the dynamical quark effects. Furthermore, in order to mimic the behavior of quark-gluon plasma it is desirable to consider geometries with appropriate factors that are dual to unquenched flavor matter at a finite temperature, see for instance \cite{Bigazzi:2009bk, Conde:2016hbg} for a related treatment in the context of the D3-D7 and D3-D5 setups, respectively.   

To consider the backreaction of the flavor in the gravity side, one has to consider the supergravity equations of motion in the presence of sources. These D-brane sources have Dirac $\delta$-functions with support on the locus of the branes. Solving the supergravity equations with point sources is challenging, and in order to circumvent these complications a standard approach has been to consider a continuous distribution of D-branes, such that no $\delta$-functions appear. This smearing approach was originally developed in \cite{Bigazzi:2005md} in the context of non-critical holography. The substitution of localized sources with a continuous distribution is valid in the limit of the large number of flavors, and hence we want both $N_c, N_f$ to be large, while their ratio, $\tfrac{N_f}{N_c}$, remains fixed to a finite value. This is the so-called Veneziano limit \cite{Veneziano:1976wm}. 

The smearing of the D-brane sources has been applied very successfully in various supergravity backgrounds to account for the flavor backreaction, see \cite{Casero:2006pt, Benini:2006hh, Benini:2007gx, Bigazzi:2009bk, Conde:2011sw, Conde:2011rg, Conde:2016hbg, Penin:2017lqt, Jokela:2019tsb, Hoyos:2020zeg, Arean:2008az, Hoyos:2021vhl}, as well as the excellent review \cite{Nunez:2010sf} for a more elaborate discussion on the smearing technique and more developments in different settings.  

There is, however, a price to pay for the above simplification. This is the fact that the dual gauge theory gets modified. The superposition of differently oriented D-branes is, essentially, a modification of the R-symmetry. In addition to that, the smeared flavor branes are not coincident, which reduces the flavor symmetry from $U(N_f)$ down to $U(1)^{N_f}$. Another important aspect is that the solutions with smeared D-branes, typically possess less amount of supersymmetry compared to solutions with localized sources. Some supersymmetries remain which helps not only to find solutions, but they can even be analytic.

Another very interesting direction that has been taken towards more general gauge/gravity duals is to consider replacing the five-dimensional internal space of the original $AdS_5 \times S^5$ background geometry by a more general five-dimensional Sasaki-Einstein manifold. This leads to the description of field theories with reduced amount of supersymmetry and falls in line with the general efforts of the community to construct and examine models that resemble much more realistic vector-like gauge theories that appear in nature than the $\mathcal{N}=4$ SYM. We will generically denote by $\mathcal{M}^{5}$ the Sasaki-Einstein space. It is known that in this context we have a duality between type IIB string theory on the $AdS_5 \times \mathcal{M}^{5}$ spacetime and a quiver gauge theory on the boundary \cite{Gubser:1998vd}. For the readers who want to get familiar with Sasaki-Einstein manifolds we suggest the review \cite{Sparks:2010sn}.

By a specific choice for the five-dimensional internal manifold $\mathcal{M}^{5} = T^{1,1}$, one arrives at the Klebanov-Witten model \cite{Klebanov:1998hh} which was the first one to be studied. However, one has derived more general classes of such five-dimensional manifolds which are characterized by either two (cohomogeneity one) or three indices (cohomogeneity two) and are denoted by $Y^{p,q}$ \cite{Gauntlett:2004yd} and $L^{a,b,c}$ \cite{Cvetic:2005ft, Martelli:2005wy} respectively. These spaces share a similar feature, namely, they all possess a base topology that is $S^2 \times S^3$. Interestingly, the metrics are explicitly known. 

The holographic field theory descriptions for the two different aforementioned families of Sasaki-Einstein manifolds are known. The dual field theory description associated to the case of the $Y^{p,q}$ manifolds was obtained in \cite{Benvenuti:2004dy, Martelli:2004wu}. The holographic gauge theory has also been worked out for the case of the $L^{a,b,c}$ manifolds \cite{Benvenuti:2005ja, Butti:2005sw,Franco:2005sm}. It is worth mentioning, that both the $T^{1,1}$ and $Y^{p,q}$ manifolds can be obtained as special cases of the general $L^{a,b,c}$ spaces. There exists, also, previous work on studying string solutions in the $Y^{p,q}$ and $L^{a,b,c}$ manifolds which can be found in \cite{Giataganas:2009dr}. There special emphasis is given on BPS configurations. Different supersymmetric D-brane embeddings in the $Y^{p,q}$ models have been studied in \cite{Canoura:2005uz}, while the authors of \cite{Canoura:2006es} focused on supersymmetric probe branes in $AdS_5 \times L^{a,b,c}$. Perhaps surprising is the fact that except for the four-dimensional $\mathcal{N}=4$ SYM associated with the $AdS_5 \times S^5$ background, none of the other field theories are integrable \cite{Basu:2011di, Basu:2011fw, Rigatos:2020hlq}. 

In this work we combine the directions described above, namely we consider backgrounds of the general form $AdS_5 \times \mathcal{M}^5$, with $\mathcal{M}^5$ being a general five-dimensional Sasaki-Einstein manifold. We consider dynamical effects on quarks due to the presence of flavor D$3'$-branes in the background. We employ the smearing approach in our studies and we derive the associated string backgrounds. 

The structure of this work is the following: in \cref{sec: sugra_setup} we describe our supergravity setup, derive the non-trivial spin-connection components, and determine the associated Killing spinors by solving the type IIB supersymmetry variations. In \cref{sec: integrating_BPS} we discuss the integration of the system of BPS equations obtained from the supersymmetry analysis. We are able to reduce the problem to that of the integration of a single second order differential equation. We discuss its solutions in two separate cases: when the quarks are massless and when the number of massive quarks is small in comparison to colors. We summarize our findings and discuss interesting future directions in \cref{sec: epilogue}. We supplement our work with two appendices. \cref{app: special_SE5} contains a discussion on various Sasaki-Einstein spaces and the associated field theories and finally, in \cref{app: metrics} we provide the explicit metric functions which are obtained after expanding in the number of flavors.

\newsec{The supergravity setup}[sec: sugra_setup]

For illustrational purposes and to have a better grasp on our setup, we schematically depict the way that the color and flavor branes are arranged in the ten-dimensional spacetime, see \cref{table: brane_scan}.
\begin{table}[H]
\begin{center}
\begin{tabular}{ |c|c|c|c|c|c|c|c|c|c|c|c|}
 \hline
 &&&&&&&&&&\\[-0.95em] 
   						& $x^0$ & $x^1$ & $x^2$ 		& $x^3$ 	& $x^4$ 		& $x^5$ 		& $x^6$ 	& $x^7$ 	& $x^8$ 	& $x^9$				\\ 
 \hline
 color D$3$ 			& --- 	& --- 	& --- 			& --- 		& $\bullet$ 	& $\bullet$ 	& $\bullet$ & $\bullet$ & $\bullet$ & $\bullet$			\\ 
 \hline
 flavor D$3'$ 			& --- 	& --- 	& $\bullet$ 	& $\bullet$ & --- 			& --- 			& $\bullet$	& $\bullet$ & $\bullet$ & $\bullet$			\\
 \hline
\end{tabular}
\caption{The supersymmetric D$3$/D$3'$ brane setup. In the above notation --- stands for a brane which is extended along that particular direction, while $\bullet$ means that the coordinate is transverse to that brane.}
\label{table: brane_scan}
\end{center}
\end{table}

\subsec{Unquenching --- Ansatz and setup}

The Ansatz for the metric is given by 
\eqna{
ds^2	&= 	h^{-1/2} \left[ -(dx^0)^2 + (dx^1)^2 + e^{m}\left( (dx^2)^2 + (dx^3)^2 \right)\right] + h^{1/2} \left( dr^2 + ds^2_5 \right)	\,		,  \\
ds^2_5	&=	 e^{2g} ds^2_{\mathrm{KE}} + e^{2f} (d\tau + A)^2 		\,		,
}[eq: ansatz_metric]
and we will work with this general way of expressing the internal manifold. In our Ansatz the dilaton remains constant, as will be shown in \cref{sec: sltn_bps_system}. We want to mention that even if one considers two different warp factors in \cref{eq: ansatz_metric} for the defect directions, $x^2$ and $x^3$, the solution of the gravitino variation yields that these have to be equal. This we checked explicitly and hence there is no loss of generality with our choice for the metric. In \cref{app: special_SE5} we discuss the specific choices $\mathcal{M}^5=\{S^5, T^{1,1}, Y^{p,q}, L^{a,b,c}\}$ for the internal part of the geometry and provide the basic expressions that one can use in the general relations we derive here. 

The above type IIB supergravity background that is depicted schematically by the array in \cref{table: brane_scan} contains a closed, self-dual five-form flux in the R-R sector, due to the stack of the background D$3$-branes. We can, thus, write 
\eqn{
F_{(5)} = I(r) (1 + \star) dx^0 \wedge dx^1 \wedge dx^2 \wedge dx^3 \wedge dr     \,          .
}[eq: flux_rr_01_color]
We can use the Bianchi identity that flux above obeys, $dF_{(5)}=0$, to determine the relation of $I(r)$ and the functions that we have used in the metric. We obtain 
\eqn{
I(r) h^2 e^{-2m} e^{4g+f} = \mathcal{I}     \,          ,
}[eq: flux_rr_02_color]
where in the above $\mathcal{I}$ is just a constant that can be related to the number of colors, $N_c$, via the flux quantization condition. Explicitly, the  relation is given by 
\eqn{
\mathcal{I} = \frac{(2 \pi)^4 g_s {\alpha^{\prime}}^2 N_c}{\mathrm{vol}_{\mathcal{M}^5}}        \,      .
}[eq: flux_rr_03_color]
where $\mathrm{vol}_{\mathcal{M}^5}$ denotes the volume of the $\mathcal{M}_5$ manifold. We can write the metric given by \cref{eq: ansatz_metric} in the orthonormal frame as
\eqn{
ds^2 = G_{MN} dx^M dx^N = \eta_{AB} E^{A} E^{B} = \eta_{AB} e^{A}{}_{M} e^{B}{}_{N} dx^M dx^N		\,			,
}[eq: ortho_frame]
with the explicit expressions for the one-form basis being given by: 
\begin{alignat}{2}
E^{\mu}			&=	h^{-1/4} dx^{\mu}								\,		,		\qquad		&&\mu=\{0,1\}					\,		,		\label{eq: one_form_basis_01}\\
E^{\bar{\mu}}	&=	h^{-1/4} e^{m} dx^{\bar{\mu}}		\,		,		\qquad		&&\bar{\mu}=\{2,3\}				\,		,		\label{eq: one_form_basis_02}\\
E^{4}			&=	h^{1/4} dr										\,		,																	\label{eq: one_form_basis_03}\\
E^{a}			&=	h^{1/4} e^{g} E^{\hat{a}}						\,		,		\qquad		&&\{a,\hat{a}\}=\{5,6,7,8\}		\,		,		\label{eq: one_form_basis_04}\\
E^{9}			&=	h^{1/4} e^{f} (d\tau + A)						\,		.																	\label{eq: one_form_basis_05}
\end{alignat}
Note that the lowercase Latin indices denote the $x^5,..,x^8$ directions of the ten-dimensional target spacetime, while hatted lowercase Latin indices specify the K\"ahler-Einstein basis. In other words, we write $ds^2_{\mathrm{KE}} = \sum\limits_{\hat{a}=5}^8 (E^{\hat{a}})^2$. For this choice of basis, the K\"ahler two-form is simply given by 
\eqn{
J = E^{\hat{5}} \wedge E^{\hat{6}} + E^{\hat{7}} \wedge E^{\hat{8}}			\,			,
}[eq: kaehler_2form]
and is related to $A$ around the fiber, $\tau$, via 
\eqn{
dA = 2 J
}[eq: J_and_A]
Since both the color degrees of freedom and the flavor degrees of freedom are associated with D3-branes, we have two contributions of the self-dual five-form in the RR-sector. The first bit comes as a contribution from the background D3-branes and is given by 
\eqn{
F^{c}_{(5)} = \mathcal{I} e^{-4g-f}h^{-5/4} \left(E^{01234} - E^{56789} \right)			\,			,
}[eq: five_form_color]
and there is another part which will account for the backreaction of the flavor, which has a non trivial pullback to the worldvolume of the D3' flavor branes  
\eqn{
C^f_4=\mathcal{Q}(r)dt\wedge dx^1 \wedge dr \wedge (d\tau+A)				\,				,			\\
}[eq: flux_001]
and it generates a five-form flux via
\eqna{
F^{f,1}_5 &= dC^f_4 			\,		,		\qquad		 F^{f,2}_5 = \star F^{f,1}_5		\,			, 		\\
F^{f}_5 &= F^{f,1}_5 + F^{f,2}_5			\,				,
}[eq: flux_dual_01]
that can be written in the one-form basis, see \cref{eq: one_form_basis_01,eq: one_form_basis_02,eq: one_form_basis_03,eq: one_form_basis_04,eq: one_form_basis_05}, explicitly as
\eqn{
F^{f}_{(5)} = -2 \mathcal{Q} e^{-2g} h^{-1/4} \left(E^{01456} + E^{01478} - E^{23569} - E^{23789} \right)			\,			,
}[eq: five_form_flavor]
having suppressed the explicit dependence of $\mathcal{Q}$ on $r$ for simplicity. We have also used the abbreviation $E^{A_1 A_2 \ldots A_n}=E^{A_1} \wedge E^{A_2} \wedge \ldots \wedge E^{A_n}$. The five-form of the setup we consider here is given by the sum of the above 
\eqn{
F_{(5)} = F^{c}_{(5)} + F^{f}_{(5)}			\,			.
}[eq: five_form_sum]
We will demand this to be self-dual in the following, but due to the explicit sources in the background, it will not be closed.

\subsec{The action and the energy-momentum tensor}

We will now look for supersymmetric solutions, so we will recast the supergravity equations of motion in a suitable form to accommodate backreacted supersymmetric sources. The total action of the system is given by:
\eqn{
S=S_{IIB}+S_{branes}				\,				,
}[eq: action_iib_branes_01]
where $S_{IIB}$ corresponds to the action of the ten-dimensional type IIB supergravity, see \cref{eq: iib_action}, and $S_{branes}$ corresponds to the action of the backreacted D$3'$ flavor branes, given by the sum of the Dirac-Born-Infeld (DBI) and Wess-Zumino (WZ) actions. The $N_f$ flavor branes of our setup act as sources of the RR five-form $F_5$. The D$3'$ flavor branes couple naturally to the RR four-form potential $C_{4}$ through the WZ term of the worldvolume action, given by:
\eqn{
S_{WZ}=T_3 \sum^{N_f}\int_{\mathcal{M}_4}\hat{C}_{(4)}				\,				,
}[eq: action_iib_branes_02]
where the hat over $C_{(4)}$ denotes its pullback to the worldvolume of the D$3'$ flavor brane and $T_3$ is the tension of the D$3'$ brane $(1/T_3=(2\pi)^3g_s (\alpha')^3)$. We will be working in the smearing approach, valid for large $N_f$, in which we substitute the discrete distribution of flavor branes by a continuous distribution with the appropriate normalization. The smearing approach amounts to perform the substitution:
\eqn{
\sum^{N_f}\int_{\mathcal{M}_4} \hat{C}_{(4)} \Longrightarrow \int_{\mathcal{M}_{10}} \Omega \wedge C_{(4)}			\,				,
}[eq: action_iib_branes_03]
where $\Omega$ is a six-form (the smearing form) with components orthogonal to the worldvolume of the flavor branes. The coupling of the flavor branes to $C_{(4)}$ induces the violation of the Bianchi identity for $F_5$, which has now a source proportional to $\Omega$. This modification can be obtained by solving the equation of motion for the $C_{(4)}$ potential which will give us now:
\eqn{
dF_5=2\kappa_{10}^2T_3 \Omega			\,				,
}[eq: violation_bianchi] 
where $2\kappa_{10}^2=(2\pi)^7g_s^2(\alpha')^4$. The DBI action can now be written in terms of the $\Omega$ form, via:
\eqn{
S_{DBI}=-T_3 \int d^{10}x \sqrt{-G_{10}}\sum_i |\Omega^{(i)}|
}[eq: action_iib_branes_04]
with $|\Omega^{(i)}|=\sqrt{\frac{1}{6!}\Omega_{ABCDEF}^{(i)}\Omega^{(i)}_{MNPQRS}G^{AM}G^{BN}G^{CP}G^{DQ}G^{ER}G^{FS}}$.

With these modifications, the equations of motion for supergravity plus sources are given by
\eqn{
\mathcal{R}_{AB}-\frac{1}{2}\eta_{AB}R-\frac{1}{2}\frac{1}{240}\left(5F_{ACDEF}F_B^{\ CDEF}-\frac{1}{2}F_{(5)}^2\right)-T^{branes}_{AB}=0 \ ,
}[eq: einstein_eom_01]
where the energy-momentum tensor is given by:
\eqna{
\frac{T^{branes}_{AB}}{-2\kappa_{10}^2T_3} = \frac{1}{2}\eta_{AB}&\sum_i |\Omega^{(i)}|-\\
&\sum_i \frac{1}{|\Omega^{(i)}|}\frac{1}{5!} (\Omega^{(i)})_{ACDEFG}(\Omega^{(i)})_{BMNPQR}\ \eta^{CM}\eta^{DN}\eta^{EP}\eta^{FQ}\eta^{GR}		\,		,	
}[eq: energy_momentum_01]

\subsec{The type IIB supersymmetry variations and Killing spinors}

To obtain supersymmetric solutions, we will first determine the Killing spinors, which amounts to solve the equations obtained from imposing the vanishing of the supersymmetric variations of dilatino and gravitino. The type IIB theory is a chiral theory, with two spinors of the same chirality and possesses $\mathcal{N}=2$ supersymmetry. The bosonic content of the theory consists of the metric, $G_{MN}$, the dilaton, $\Phi$, a two-form in the NS-NS sector, $B_{(2)}$, and a zero-,two-, and four-form in the R-R sector, $A_{(0)}$, $A_{(2)}$, and $A_{(4)}$. The type IIB supergravity action in the string frame is given by 
\eqna{
S=&\frac{1}{2 \kappa^2_{10}} \int d^{10}x \sqrt{-G} \left(e^{-2\phi} \left(R + 4 \partial_M \Phi \partial^M \Phi - \frac{1}{12} H^2_{(3)} \right) -\frac{1}{2} F^2_{(1)} - \frac{1}{12} F^2_{(3)} - \frac{1}{480} F^2_{(5)} \right)			\,			\\
&+\frac{1}{4 \kappa^2_{10}} \int dA_{(2)} \wedge H_{(3)} \wedge \left(A_{(4)} + \frac{1}{2} B_{(2)} \wedge A_{(2)} \right)		\,			,
}[eq: iib_action]
with 
\eqn{
H_{(3)} = dB_{(2)}	\,	,	\quad	F_{(1)} = dA_{(0)}	\,	,	\quad F_{(3)} = dA_{(2)} + C_{(0)}H_{(3)}	\,	,	\quad F_{(5)} = dA_{(4)} + H_{(3)} \wedge A_{(2)}	\,	.
}[eq: aux_iib_relations]
The equations of motion derived from the supergravity action have to be supplemented by the self-duality condition on the five-form; $F_{(5)} = \star F_{(5)}$. The spinor, $\epsilon$, that parameterizes the supersymmetry transformations consists of two Majorana-Weyl spinors and has a well-defined chirality. 

The supersymmetry transformations for the dilatino and the gravitino are given by
\eqna{
\delta \lambda	&=	\left[\frac{1}{2}\slashed{\partial}\Phi + \frac{1}{4\cdot 3!} \slashed{H}_{(3)} \tau_3 - \frac{e^{\Phi}}{2} \slashed{F}_{(1)} (i \tau_2) - \frac{e^{\Phi}}{4 \cdot 3!} \slashed{F}_{(3)} \tau_1 \right]\epsilon	\,	,	\\
\delta \psi_{M}	&=	\left[\nabla_M + \frac{1}{4\cdot 2!} H_{MNR}\Gamma^{NR} \tau_3 + \frac{e^{\Phi}}{8} \left( \slashed{F}_{(1)} (i \tau_2) + \frac{1}{3!} \slashed{F}_{(3)} \tau_1 + \frac{1}{2\cdot 5!} \slashed{F}_{(5)} (i \tau_2) \right) \Gamma_M \right]\epsilon	\,	,
}[eq: app_susy_01]
where we have used a shorthand notation $\slashed{\mathcal{X}}_{(M)} = \Gamma^{M_1 M_2 ... M_N} \mathcal{X}_{M_1 M_2 ... M_N}$ and $\tau_i$ are the standard Pauli matrices.

For the background we consider here, the type IIB supersymmetry variations read
\eqna{
\delta \lambda		&=		\frac{1}{2} \Gamma^{M} \left( \partial_{M} \Phi \right)\epsilon																								\,		,			\\
\delta \psi_{M}		&=		\left[ \nabla_{M} + \frac{e^{\Phi}}{1920} F_{ABCDE} \Gamma^{ABCDE} \Gamma_{M} \left(i \tau_2 \right)	\right] \epsilon		\,		.
}[eq: susy_vars_iib]
In order to analyze the supersymmetry variations and determine the Killing spinors, we demand that the variations above vanish.

\subsec{Equations for the BPS system}[sec: sltn_bps_system]

Since the covariant derivative appears in the gravitino variation, \cref{eq: susy_vars_iib}, we need to determine the non-vanishing components of the spin-connection. In order to compute these non-vanishing spin-connection components, we will use the Cartan structure equation. For torsion-free theories, it is given by 
\eqn{
dE^{A} + \omega^{A}{}_{B} \wedge E^{B} = 0		\,		.
}[eq: cartan_eqn]
A straightforward calculation yields
\begin{alignat}{2}
\omega^{\mu}{}_{4}						&=	- \frac{1}{4} h^{-5/4} h^{\prime} E^{\mu}																				\,	,		\qquad		&&\mu=\{0,1\}				\,		,		\\
\omega^{\bar{\mu}}{}_{4}				&=	\left(-\frac{1}{4} h^{-5/4} h^{\prime} + h^{-1/4} m^{\prime} \right) E^{\bar{\mu}}							\,	,		\qquad		&&\bar{\mu}=\{2,3\}		\,		,		\\
\omega^{a}{}_{4}						&=	\left( \frac{1}{4} \frac{h^{\prime}}{h} + g^{\prime} \right)h^{-1/4}E^{a}												\,	,		\qquad		&&a=\{5,6,7,8\}		\,		,		\\
\omega^{9}{}_{4}						&=	\left( \frac{1}{4} \frac{h^{\prime}}{h} + f^{\prime} \right)h^{-1/4}E^{9}												\,		,		\\
\omega^{9}{}_{a}						&=	h^{-1/4} e^{f-2g}J_{\hat{a} \hat{b}}	E^{b}																			\,	,		\qquad		&&\{a,b\}=\{5,6,7,8\}		\,		,		\\
\omega^{a}{}_{b}						&=	\omega^{\hat{a}}{}_{\hat{b}} - h^{-1/4} e^{f-2g} J^{\hat{a}}{}_{\hat{b}} E^{9}											\,	,		\qquad		&&\{a,b\}=\{5,6,7,8\}		\,		,		
\end{alignat}\label{eq: spin_conn_all}
where we have used the abbreviation $\partial_r \equiv {}^{\prime}$.

The vanishing of the dilatino variation, $\delta \lambda = 0$, can be readily solved by 
\eqn{
\Phi = \mathrm{constant}		\,		.
}[eq: dilatino_sltn] 
It is worth commenting on this result. In the D$3$-background solution of type IIB theory, the dilaton is just a constant. In other backreacted systems based on smearing flavor branes on the D$3$-background the dilaton attains a non-trivial profile \cite{Bigazzi:2009bk, Conde:2016hbg}. In our case, however, the dilaton remains intact. The associated flux is a new contribution to the five-form and no new terms in the dilatino variation arise.  This is an important result as it implies that the classical supergravity solution that we obtained is a trustworthy result for any number of flavors.

We now study the equation coming from the gravitino variation, $\delta \psi_{M}=0$, and we consider first the Minkowski components of the equation. In what follows, all indices are flat, i.e they are defined in the one-form basis. We start by considering the $M=0$ component and we have 
\eqna{
\bigg\{-\frac{1}{8} h^{-5/4} h^{\prime} \Gamma_{0 4} + &\frac{i \tau_2}{16} \left(\mathcal{I} e^{-4g-f} h^{-5/4} \left(\Gamma^{01234}-\Gamma^{56789}\right) \right. \\
&\left. - 2 \mathcal{Q} e^{-2g} h^{-1/4} \left(\Gamma^{01456}+\Gamma^{01478}-\Gamma^{23569}-\Gamma^{23789}\right) \right) \Gamma_{0} \bigg\} \epsilon = 0		\,			.
}[eq: gravitino_mu_aux_01]
From the ten-dimensional chirality condition in type IIB supergravity on the spinor
\eqn{
\Gamma^{0123456789} \epsilon = - \epsilon						\,			,
}[eq: projection_01]
we obtain that 
\eqn{
\Gamma^{01234} \epsilon = \Gamma^{56789} \epsilon				\,			.
}[eq: projection_02]
In addition to the above, we  will impose the projection
\eqn{
\Gamma_{0123} \left(i \tau_2 \right) \epsilon = \epsilon		\,			,
}[eq: projection_03]
which corresponds to placing the stack of (color) D3-branes along the $\{x^0,x^1,x^2,x^3\}$. Furthermore, we will impose the K\"ahler condition
\eqn{
\Gamma^{56} \epsilon = \Gamma^{78} \epsilon		\,			.
}[eq: projection_04]
Using the above, \cref{eq: gravitino_mu_aux_01} becomes
\eqn{
\bigg\{-\frac{1}{8} \frac{h^{\prime}}{h^{5/4}} - \frac{\mathcal{I}}{8h^{5/4}}e^{-4g-f} - 2 \mathcal{Q} e^{-2g} h^{-1/4} (i \tau_2) \left(\Gamma^{0156} + \Gamma^{234569} \right) \bigg\}\epsilon = 0		\,				.
}[eq: gravitino_mu_aux_02]
Some straightforward algebra with the term involving $\Gamma$-matrices yields
\eqn{
(i \tau_2) \left(\Gamma^{0156} + \Gamma^{234569} \right) \epsilon = \Gamma^{0156} \left(i \tau_2 - \Gamma^{49} \right)\epsilon		\,				.
}[eq: gravitino_mu_aux_03]
We impose the following projections
\eqn{
\Gamma^{49} \epsilon = - (i \tau_2) \epsilon			\,			,		\qquad		\Gamma^{015649} \epsilon = -\epsilon			\,			.
}[eq: gravitino_mu_aux_04]
Imposing the above, \cref{eq: gravitino_mu_aux_02} becomes
\eqn{
h^{\prime} + \mathcal{I} e^{-4g-f} + 4 \mathcal{Q} e^{-2g} h = 0 		\,			.
}[eq: gravitino_mu_aux_05]
Examining the $M=1$ component of the gravitino equation produces the same equation as \cref{eq: gravitino_mu_aux_05}. 

Next, we examine the $M=2$ (or equivalently $M=3$) components of the gravitino variation. We obtain
\eqna{
\bigg\{&-\left( \frac{1}{8} h^{-5/4} h^{\prime} - \frac{1}{2} m^{\prime} \right)\Gamma_{\bar{\mu} 4} + \frac{i \tau_2}{16} \left(\mathcal{I} e^{-4g-f} h^{-5/4} \left(\Gamma^{01234}-\Gamma^{56789}\right) \right. \\
&\left. - 2 \mathcal{Q} e^{-2g} h^{-1/4} \left(\Gamma^{01456}+\Gamma^{01478}-\Gamma^{23569}-\Gamma^{23789}\right) \right) \Gamma_{\bar{\mu}} \bigg\} \epsilon = 0		\,			.
}[eq: gravitino_mu_bar_aux_01]
We can use the same projections as before, see \cref{eq: projection_01,eq: projection_02,eq: projection_03,eq: gravitino_mu_aux_04}, and we arrive at 
\eqn{
\frac{h^{\prime}}{h} - 4~m^{\prime} + \frac{\mathcal{I}}{h} e^{-4g-f} - 4 \mathcal{Q} e^{-2g} = 0		\,			.
}[eq: gravitino_mu_bar_aux_02]
Finally, we can use the expression for $h^{\prime}$ from \cref{eq: gravitino_mu_aux_05} and obtain 
\eqn{
m^{\prime} = - 2 \mathcal{Q} e^{-2g}				\,		.
}[eq: gravitino_mu_bar_aux_05]
We proceed to examine the radial component of the gravitino equation. We obtain 
\eqna{
\bigg\{h^{-1/4} \partial_4 &+ \frac{i \tau_2}{16} \left(\mathcal{I} e^{-4g-f} h^{-5/4} \left(\Gamma^{01234}-\Gamma^{56789}\right) \right. \\
&\left. - 2 \mathcal{Q} e^{-2g} h^{-1/4} \left(\Gamma^{01456}+\Gamma^{01478}-\Gamma^{23569}-\Gamma^{23789}\right) \right) \Gamma_{4} \bigg\} \epsilon = 0		\,			,
}[eq: gravitino_r_aux_01]
and using the projections written previously we can simplify the above to the following: 
\eqn{
\bigg\{ \partial_4 - \frac{1}{8} \frac{\mathcal{I}}{h} e^{-4g-f} - \frac{1}{2} \mathcal{Q} e^{-2g} \bigg\} \epsilon = 0		\,			.
}[eq: gravitino_r_aux_02]
We can use \cref{eq: gravitino_mu_aux_05} and re-write \cref{eq: gravitino_r_aux_02} as 
\eqn{
\partial_4 \epsilon = - \frac{h^{\prime}}{8h} \epsilon		\,			.
}[eq: gravitino_r_aux_03]
The above equation has an obvious solution, which is given by 
\eqn{
\epsilon = h^{-1/8} \eta		\,			,
}[eq: gravitino_r_aux_04]
with $\eta$ being a spinor that does not depend on the radial coordinate.
The final step is to analyze the equations associated with the internal part of the geometry. 

Let us now study the directions along the K\"ahler-Einstein basis. We consider as an example the case $M=5$ (the $M=\{6,7,8\}$ components lead to the same equation by symmetry). We obtain 
\eqna{
\bigg\{e^{-g}\Gamma_{54}\left(\hat{D}_5 - A_5 \partial_{\tau}\right) + \frac{1}{2} e^{f-2g} \Gamma_{9456} - \frac{1}{2}\left(\frac{h^{\prime}}{4h} + g^{\prime} \right) - \frac{1}{8} \frac{\mathcal{I}}{h}e^{-4g-f} \bigg\}\epsilon = 0		\,			,
}[eq: gravitino_ke_basis_aux_01]
where in the above $\hat{D}$ is the covariant derivative along the K\"ahler-Einstein basis and $A$ is the one-form potential, see \cref{eq: J_and_A}. It is easy to see that from the previously used projections we can derive 
\eqn{
\Gamma_{9456} \epsilon = \epsilon			\,		.
}[eq: new_projection]
In order to proceed, we will use the fact that in any K\"ahler-Einstein space there exists a covariantly constant spinor which satisfies the following
\eqn{
\hat{D}_i \epsilon = \frac{3}{2} \Gamma_{56} A_i \epsilon = - \frac{3}{2} (i \tau_2) A_i \epsilon		\,			,
}[eq: gravitino_ke_basis_aux_02]
with $i$ labeling any of the directions along the K\"ahler-Einstein basis; $i=\{5,6,7,8\}$. In the basis of one-forms of the K\"ahler-Einstein space that we are using, the Killing spinor can be taken to be independent of the $\{5,6,7,8\}$ coordinates, and it has only a dependence on the fiber, $\tau$. Hence, we have:
\eqn{
\hat{D}_i \epsilon - A_i \partial_{\tau} \epsilon = - A_i \left(\partial_{\tau} \epsilon + \frac{3}{2} (i \tau_2) \epsilon \right)			\,				, 
}[eq: gravitino_ke_basis_aux_03]
which we can make it vanish with the following dependence of the Killing spinor on the fiber, $\tau$, 
\eqn{
\partial_{\tau} \epsilon = - \frac{3}{2} (i \tau_2) \epsilon = \frac{3}{2} \Gamma_{56} \epsilon		\,			.
}[eq: gravitino_ke_basis_aux_04]
From the above, \cref{eq: gravitino_ke_basis_aux_01} becomes 
\eqn{
-\left(g^{\prime}+\frac{h^{\prime}}{4h}\right) + e^{f-2g} - \frac{1}{4} \frac{\mathcal{I}}{h} e^{-4g-f}		=		0		\,		.
}[eq: gravitino_ke_basis_aux_05]
We can use the result \cref{eq: gravitino_mu_aux_05} and re-write the above as
\eqn{
g^{\prime} = e^{f-2g} + \mathcal{Q}e^{-2g}		\,		.
}[eq: gravitino_ke_basis_aux_06]
Finally, we want to consider the fiber, given by $M=9$. The gravitino equation in this direction becomes 
\eqna{
\bigg\{\Gamma_{94}\left(e^{-f}\partial_{\tau}-e^{f-2g}\Gamma_{56}\right)-\frac{1}{2}\left(\frac{h^{\prime}}{4h}+f^{\prime}\right)-\frac{1}{8}\frac{\mathcal{I}}{h}e^{-4g-f}-\frac{1}{2}\mathcal{Q}e^{-2g}\bigg\}\epsilon		=		0 		\,			,
}[eq: gravitino_fiber_aux_01]
and by using that 
\eqn{
\partial_{\tau} \epsilon = \frac{3}{2}\Gamma_{56}\epsilon			\,				,
}[eq: gravitino_fiber_aux_02]
we obtain
\eqn{
\frac{3}{2}e^{-f} - 2 e^{f-2g} - \frac{1}{2}\left(\frac{h^{\prime}}{4h}+f^{\prime}\right) - \frac{1}{8}\frac{\mathcal{I}}{h}e^{-4g-f} - \frac{1}{2}\mathcal{Q}e^{-2g} = 0		\,			.
}[eq: gravitino_fiber_aux_03] 
Finally, we can use \cref{eq: gravitino_mu_aux_05} and we can write the above equation as 
\eqn{
3e^{-f}-2e^{f-2g} - f^{\prime} = 0		\,			.
}[eq: gravitino_fiber_aux_04]

Finally, some straightforward algebra reveals the following relation
\eqn{
\mathcal{I} = -h^{\prime} e^{4g+f} + \mathcal{Q} h e^{2g+f}     \,          ,
}[eq: mathcalI_metric_01]
which can be used with \cref{eq: flux_rr_02_color} to fully specify our setup.

\subsubsecstar{Summary of equations and projections}

To summarize, after imposing the vanishing of the supersymmetry variations, we have obtained the following system of first-order differential equations:
\eqna{
\Phi^{\prime}			&=	0											\,		,				\\
h^{\prime}				&=	-\mathcal{I} e^{-4g-f} + 4\mathcal{Q}~h~e^{-2g}		\,		,				\\
g^{\prime}				&=	-\mathcal{Q}e^{-2g} + e^{f-2g}				\,		,				\\
f^{\prime}				&=	3e^{-f}-2e^{f-2g}							\,		,				\\
m'\equiv 			m_i &= 2\mathcal{Q}e^{-2g}							\,		,				\qquad			i=\{2,3\}	\\
\partial_4 \epsilon		&=	- \frac{h^{\prime}}{8h} \epsilon			\,		.
}[eq: bps_summary]
where the last equation above determines the radial dependence of the Killing spinor. The projections we used are:
\eqna{
\Gamma^{0123456789} \epsilon					&=			- \epsilon						\,			,			\\
\Gamma^{01234} \epsilon 						&= 			\Gamma^{56789} \epsilon			\,			,			\\
\Gamma_{0123} \left(i \tau_2 \right) \epsilon 	&= 			\epsilon						\,			,			\\
\Gamma^{56} \epsilon 							&= 			\Gamma^{78} \epsilon			\,			,			\\
\Gamma^{49} \epsilon 							&= 			- (i \tau_2) \epsilon			\,			,			\\
\Gamma^{015649} \epsilon 						&= 			-\epsilon						\,			.
}[eq: projections_summary]

One can immediately see that this BPS system solves the second order equations of motion, and we can now obtain the energy-momentum tensor associated to the brane, and the smearing form, which are given by: 
\eqna{
&\Omega=\Omega^{(1)}+\Omega^{(2)}+\Omega^{(3)}\\
&\Omega^{(1)}=h^{-\frac{1}{2}}
\frac{1}{2\pi^4}\mathcal{Q}e^{f-4g} E^{235678}																					\,				,		\\
&\Omega^{(2)}=h^{-\frac{1}{2}}
\frac{1}{8\pi^4}\left(-2e^{f-4g}\mathcal{Q}+3e^{-f-2g}\mathcal{Q}-4e^{-4g}\mathcal{Q}^2+e^{-2g}\mathcal{Q}^{\prime}\right)  E^{234569}		\,				,		\\
&\Omega^{(3)}=h^{-\frac{1}{2}}
\frac{1}{8\pi^4}\left(-2e^{f-4g}\mathcal{Q}+3e^{-f-2g}\mathcal{Q}-4e^{-4g}\mathcal{Q}^2+e^{-2g}\mathcal{Q}^{\prime}\right)  E^{234789}		\,				.
}[eq: omega_defs_01]

and

\eqna{
&T^{branes}_{\mu \nu}=\eta_{\mu \nu}h^{-\frac{1}{2}}\left(-6\mathcal{Q} e^{-f-2g}+8\mathcal{Q}^2e^{-4g}-2e^{-2g}\mathcal{Q}^{\prime}\right)		\,		, 	\qquad	 \{\mu, \nu\}=\{0,1\}		\,			,		\\
&T^{branes}_{22}=T^{branes}_{33}=0																												\,		,								\\
&T^{branes}_{44}=-4h^{-\frac{1}{2}}\mathcal{Q}e^{f-4g}																							\,		,								\\
&T^{branes}_{ii}=h^{-\frac{1}{2}}\left(2e^{f-4g}\mathcal{Q}-3e^{-f-2g}\mathcal{Q}+4e^{-4g}\mathcal{Q}^2-e^{-2g}\mathcal{Q}^{\prime}\right)		\,		,	\qquad 	i=\{5,6,7,8\}				\,			,		\\
&T^{branes}_{99}=-h^{-\frac{1}{2}}4e^{f-4g}\mathcal{Q}																							\,		.
}[eq: energy_momentum_02]

\newsec{Analysis of the equations}[sec: integrating_BPS]
In this section we will briefly discuss solutions to the BPS system that we have previously derived. 
We want to study profiles $\mathcal{Q}(r)$ that are suitable to represent flavor degrees of freedom. In principle these would be dictated directly by the kappa symmetry if one were to make contact with microscopic embeddings with which one performs the smearing procedure. This procedure depends on the internal geometry, and since we aim to be general, we do not specify $\mathcal{Q}(r)$ in great detail in the following.

\subsec{Small flavor expansion}[sec: flavorexpansion]

First of all, given $\mathcal{Q}(r)$, one can directly solve the differential equations \cref{eq: bps_summary} using numerical methods; see e.g.~\cite{Bea:2013jxa}. As mentioned above, for this purpose one needs to specify the internal geometry and perform the kappa symmetry analysis which yields the corresponding profile $\mathcal{Q}(r)$. However, much can be said without doing this analysis and indeed, there are good reasons to relaxing strict requirements on matching to microscopics \cite{Hoyos:2021vhl}.

Second, to make the analysis more compact, we can actually combine several of the above differential equations. To this end, let us combine the BPS equations for $g$ and $f$, see \cref{eq: bps_summary}, into a single second order differential equation. We have
\eqn{
g=\frac{f}{2}-\frac{1}{2}\log\left(-\frac{3}{2}e^{-f}-\frac{f^{\prime}}{2} \right)							\,				,
}[eq: int_bps_01]
and, thus, we get the following expression for the $f$ function
\eqn{
f^{\prime \prime}-\left(9e^{-2f}+9e^{-3f}\mathcal{Q}+(-12e^{-f}-6e^{-2f}\mathcal{Q})f'+(2+e^{-f}\mathcal{Q}){f^{\prime}}^2\right)=0			\,				,
}[eq: int_bps_02]
In this and the following subsections we will aim to solve this differential equation.

Let us start by solving the BPS system perturbatively in the small flavor limit. By defining $\epsilon\equiv N_f/N_c$, such regime is attained by assuming the limit $\epsilon\ll 1$.

We start by assuming an Ansatz of the form: 
\eqna{
\mathcal{Q}		&=\epsilon~q_0				\,		\\
f				&=f_0+\epsilon~f_1			\,		\\
m				&=\epsilon~m_1	\,		\\
g				&=g_0+\epsilon~g_1			\,		\\	
h				&=h_0+\epsilon~h_1			\,		.
}[eq: int_bps_03]
We will take the solution for the $\{f_0,g_0,h_0\}$ to correspond to the AdS solution, and hence $f_0=g_0=\log r$. Using this it is quite straightforward to obtain:
\eqn{
f^{\prime \prime}_1+\frac{8}{r}f^{\prime}_1+6\frac{f_1}{r^2}-4\frac{q_0}{r^3}=0
}[eq: int_bps_04]
which admits a solution given by 
\eqn{
f_1=\frac{c_{f_1}}{r}+\frac{c_{f_2}}{r^6}+\frac{4}{5r}\int_1^r \frac{q_0(z)}{z}dz-\frac{4}{5r^6}\int_1^r z^4q_0(z)dz 		\,				,
}[eq: int_bps_05]
with which we can immediately obtain the following for $g_1$:
\eqn{
g_1=\frac{c_{f_1}}{r}-\frac{c_{f_2}}{4r^6}+\frac{4}{5r}\int_1^r \frac{q_0(z)}{z}dz+\frac{1}{5r^6}\int_1^r z^4q_0(z)dz		\,				.
}[eq: int_bps_06]
For the warp factors along the $x^2, x^3$ directions, we have
\eqn{
m_1^{\prime}+2\frac{q_0}{r^2}=0							\,				,
}[eq: int_bps_07]
which we can solve in a very straightforward manner as:
\eqn{
m_1=c_m-2\int_1^r\frac{q_0(z)}{z^2}dz				\,				.
}[eq: int_bps_08]
We are now left with a differential equation for $h_1$, which, after using the AdS value $h_0=\frac{R}{4r^4}$, is given by:
\eqna{
&0=h_1'+\frac{R}{r^6}(q_0(r)-5c_{f_1})-\frac{4R}{r^6}\int_{1}^r \frac{q_0(z)}{z}dz			\,		,
}[eq: int_bps_09]
which can be solved as 
\eqn{
h_1=\int_1^r \frac{R}{z^{6}}\left(5 c_{f_1}-q_0(z)+4\int_1^z\frac{q_0(w)}{w}dw \right) dz			\,			.
}[eq: int_bps_10]

We just obtained closed set of integrals which completely determine the perturbed metric components if the seed $q_0(r) $ (limit of $\mathcal{Q}(r)$) is given. We have required that in the flavorless limit $\epsilon\to 0$ one obtains the original $AdS_5\times \mathcal{M}^5$. We have also demanded an overall additive integration constant in \cref{eq: int_bps_10} to vanish to have a sensible warping at any nonzero number of flavors. In the next subsections we will assume some specific forms for $\mathcal{Q}(r)$, which is enough to fix all the remaining integration constants.

\subsec{Constant profile}
We start with a simple choice for the seed $q_0(r)$. Let us consider constant $q_0$, i.e.~not depending on the radial coordinate: 
\eqn{
q_0(r)= q_0         \,          .
}[eq: flavor_prof_01_cnstnt]
This choice would correspond to having massless flavor D3-branes; they span the full radial coordinate. A straightforward calculation leads to the following solution for the various functions that determine our system
\eqna{
g				&=\log r + \epsilon~\frac{1}{100r^6}\left(4q_0 r^5 + 100 c_1 r^5 + 80 q_0 r^5 \log r  \right)																				\,		,		\\
f				&=\log r + \epsilon~\left(\frac{c_1}{r}  + \frac{4q_0 (5\log r-1)}{25r} \right)																				\,		,		\\
h				&=\frac{R}{4r^4}-\epsilon~\left(\frac{q_0 R}{25 r^5}-\frac{c_1 R}{r^5}-\frac{4q_0 R\log r}{5r^5}\right)																		\,		,		\\
m_1	&=\epsilon \left(c_1 + \frac{2q_0}{r} \right)		\,		,		\\
q				&=\epsilon q_0		\,		.
}[eq: flavor_prof_11_cnstnt]

We note that there are several constants.  Notice that there are some terms that would seem to lead to curvature singularities. This is not the case, however, as it will be discussed later in \cref{sec:impli}.
\subsec{Non-trivial profile}

We proceed to examine a non-trivial profile. We consider
\eqn{
q_0(r)=\frac{r^3}{1+r^3}
}[eq: flavor_prof_01]
where the exponent could in principle be anything larger than 2 to avoid divergences, but here we choose to steer clear of clumpliness. This simple choice for the profile not only allows us to find rather simple solutions, but is also motivated by the similarity to massive profiles; see also similar considerations in \cite{Hoyos:2021vhl}. After a straightforward computation we get
\eqna{
g				&=\log r + \epsilon~\mathcal{G}																				\,		,		\\
f				&=\log r - \epsilon~\mathcal{F}																				\,		,		\\
h				&=\frac{R}{4r^4}-\epsilon~\mathcal{H}																		\,		,		\\
m_1	&=\epsilon \left(a_i-\frac{2}{3} \sqrt{3} \arctan \left(\frac{2r-1}{\sqrt{3}}\right)+\frac{2}{3}\log(r+1)-\frac{1}{3}\log(r^2-r+1)\right)		\,		,		\\
q				&=\epsilon \frac{r^3}{1+r^3}		\,		,
}[eq: flavor_prof_11]
with 
\eqna{
&\mathcal{G}		=	\frac{1}{450r^6}\left(\mathcal{W} + 120r^5 \log (r^3+1) \right)		\,	,	\\
&\mathcal{F}		=	\frac{2}{225r^6}\left(\mathcal{W} - 30r^5  \log (r^3+1) \right)		\,	,	\\
&\mathcal{H}		=	\frac{R}{30} \left(\sqrt{3}\pi-\frac{3}{r^2}-2  \sqrt{3} \arctan \left(\frac{2r-1}{\sqrt{3}}\right)- 2\log (r+1)+\right. \\
&\left.\log(r^2-r+1)+\frac{8}{r^5}\log(r^3+1) \right)		\,	,
}[eq: flavor_prof_12]
and also
\eqna{
\mathcal{W} &= 5\sqrt{3}\pi - 45r^2 + 18r^5 + 30 \sqrt{3} \arctan \left(\frac{2r-1}{\sqrt{3}}\right) - 30\log (r+1) + 15 \log(r^2-r+1)			.		
}[eq: def_of_W]
\subsec{Implications on the geometry}[sec:impli]

Let us discuss the implications of the above solutions at the level of the geometry. It is a straightforward exercise to replace our solutions for a constant profile given by \cref{eq: flavor_prof_11_cnstnt} as well as the ones we described for a specific non-trivial dependence of the holographic coordinate, \cref{eq: flavor_prof_11,eq: flavor_prof_12,eq: def_of_W}, into the expression for the geometry, \cref{eq: ansatz_metric}. 
We begin by discussing the case of a constant profile. We are working to leading order in the small flavor expansion $\epsilon\ll 1$, with $\epsilon \sim N_f$. We want to examine both the UV and IR limits of the metric. We obtain:
\eqna{
ds^2 = &\left( \frac{2r^2}{\sqrt{R}} + \mathcal{A}_1 \right) (-(dx^0)^2+(dx^1)^2) + \left( \frac{2r^2}{\sqrt{R}} + \mathcal{A}_2 \right) ((dx^2)^2+(dx^3)^2) +\\
&\left( \frac{2r^2}{\sqrt{R}} + \mathcal{A}_3 \right) dr^2 + \left( \frac{\sqrt{R}}{2} + \mathcal{A}_4 \right) ds^2_{KE} + \left( \frac{\sqrt{R}}{2} + \mathcal{A}_5 \right) (d\tau + A)^2 \,   ,   
}[eq: geometry_epsilon_01]
where the various functions are listed in \cref{app: metrics}; see \cref{eq: fnctn_aux_01}. We can work in the same vein for the profile with the non-trivial radial dependence. Again, the various functions that will appear in \cref{eq: geometry_epsilon_01} after the appropriate IR and UV expansions are listed in \cref{app: metrics} and more precisely \cref{eq: fnctn_aux_02}. 
Before discussing the contributions from the flavors, notice that with the identification 
$R \rightarrow 4 R^4_{AdS},R \rightarrow   4 R^4_{5}$ one immediately finds $ds^2 = ds^2_{AdS_5} + ds^2_{5}$ in its canonical form.

We want to examine the behavior of the non-vanishing components of the energy-momentum tensor for the two different kind of solutions that we considered above. Let us present the results up to first order in $\epsilon$. For the constant profile solutions a straightforward evaluation of \cref{eq: energy_momentum_02} on the solutions yields:
\eqna{
T^{\mathrm{branes}}_{\mu \nu}   &=      -\eta_{\mu \nu} \frac{12q_0}{r\sqrt{R}}~\epsilon      \,          ,       \\
T^{\mathrm{branes}}_{44}        &=      -\frac{8 q_0 r^2}{\sqrt{R}}~\epsilon      \,          ,       \\
T^{\mathrm{branes}}_{ii}        &=      -\frac{2q_0}{r \sqrt{R}}~\epsilon      \,          ,       \\
T^{\mathrm{branes}}_{99}        &=       -\frac{8 q_0 r^2}{\sqrt{R}}~\epsilon      \,          ,       
}[eq: stress_energy_01]
One could proceed similarly to find next-to-leading order results.

From the above, we can see that all contributions to the energy-momentum tensor are finite, and also that $T_{00} \geq 0$ as long as $q_0 \geq 0$. The condition $q_0 \geq 0$ is very reasonable as this constant is related to the number of the flavor degrees of freedom. 

Let us also spell out the contributions to the energy momentum tensor for the non-trivial profile solutions we considered in the previous section. These read
\eqna{
T^{\mathrm{branes}}_{\mu \nu}   &=      -\eta_{\mu \nu} \frac{12r^2(r^3+2)}{(r^3+1)^2\sqrt{R}}~\epsilon      \,          ,       \\
T^{\mathrm{branes}}_{44}        &=      -\frac{8 r^5}{(r^3+1)\sqrt{R}}~\epsilon      \,          ,       \\
T^{\mathrm{branes}}_{ii}        &=      -\frac{2r^2(r^3+4)}{(r^3+1)^2 \sqrt{R}}~\epsilon      \,          ,       \\
T^{\mathrm{branes}}_{99}        &=       -\frac{8 r^5}{(r^3+1)\sqrt{R}}~\epsilon      \,          .      
}[eq: stress_energy_02]
Notice that again all contributions are finite and there are no negative energy solutions since $T_{00} \geq 0$.

We want to examine, additionally, whether our solutions are smooth or they exhibit curvature singularities. To do so, we compute the Ricci and the Kretschmann scalars and the square of the Ricci tensor for the different choices of the profile functions we studied in this work. Before we present our results we remind the basic formulae that we need to use for the reader's convenience. 

We can calculate the Ricci, $\mathcal{R}_{AB}$, and Riemann, $\mathcal{R}_{ABCD}$, tensors' components in flat coordinates for our ten-dimensional metric by using\footnote{For a detailed account, we refer the reader to \cite[section 3.3]{Benini:2006hh} which discusses Sasaki-Einstein manifolds.} 
\eqna{
\mathcal{R}^{A}{}_{B} &= d\omega^{A}{}_{B} + \omega^{A}{}_{C} \wedge \omega^{C}{}_{B}     \,         ,           \\
\mathcal{R}^{A}{}_{B} &= \frac{1}{2} \mathcal{R}^{A}{}_{BCD} E^{C} \wedge E^{D}                     \,         ,
}[eq: ricci_riemann_01]
and then we need to perform the following contractions
\eqn{
\mathcal{R} = G^{MN}\mathcal{R}_{MN}        \,      ,       \qquad     \mathcal{R}^2 = \mathcal{R}^{MN}\mathcal{R}_{MN}      \,      ,       \qquad      K = \mathcal{R}^{ABCD} \mathcal{R}_{ABCD}       \,          ,
}[eq: curvature_scalars]
The resulting expressions are lengthy and hence we refrain from providing them explicitly. We instead find it more illuminating to discuss the results for the curvature invariants for the two different profiles considered in this work.

For the ten-dimensional geometry under consideration and the constant profile solutions, we have (up to leading order in the $\epsilon$-expansion): 
\eqna{
\mathcal{R}       &=  \frac{6(\mathcal{I}-R)}{R^{3/2}}             \,      ,       \\   
\mathcal{R}^2     &=  640\frac{\mathcal{I}^4}{R^5}                \,          ,       \\  
K       &=  3840\frac{\mathcal{I}^2}{R^3} - 7680\frac{\mathcal{I}^3}{R^4} + 4160\frac{\mathcal{I}^4}{R^5}       \,          ,
}[eq: sphere_curvature_02]
and for the profile with a functional dependence on the $r$-coordinate we obtain:
\eqna{
\mathcal{R}       &\sim  12.5\frac{(\mathcal{I}-R)}{R^{3/2}}             \,      ,       \\   
\mathcal{R}^2     &=  640\frac{\mathcal{I}^4}{R^5}                \,          ,       \\  
K       &=  3840\frac{\mathcal{I}^2}{R^3} - 7680\frac{\mathcal{I}^3}{R^4} + 4160\frac{\mathcal{I}^4}{R^5}       \,          ,
}[eq: sphere_curvature_03]
and we remind the reader of \cref{eq: flux_rr_02_color} which was used in order to express the curvature invariants in terms of $\mathcal{I}$ \cref{eq: flux_rr_03_color}. Since the above are constants, the metrics that are related to the solutions derived in this work are free of curvature singularities.

\newsec{Conclusions}[sec: epilogue]

In this work we considered supersymmetric solutions that are realized at the intersections of color D$3$- and flavor D$3'$-branes. The latter furnish a codimension two defect on the worldvolume of the background D$3$-branes. In theories that accommodate fundamental degrees of freedom, they are confined on the defect surface and all composite states propagate in the two-dimensional subspace. The backgrounds we derived solve consistently the equations of motion of type IIB supergravity in the presence of sources. These sources correspond to a smearing of the flavor D-branes.

The smeared flavor D$3'$-branes are placed on the tip of a Calabi-Yau cone that can be constructed with the use of a five-dimensional Sasaki-Einstein space. This generic case we examined encodes the explicit constructions of $S^5, T^{1,1}, Y^{p,q},$ and $L^{a,b,c}$, which we describe in \cref{app: special_SE5}. We derived a system of first-order differential equations, the BPS conditions, that we were able to integrate.  This achievement is an important stepping stone for further explorations, which we will discuss next.

Below we will give our undoubtebly biased list of interesting future research directions:

We have restricted our attention to the study of the geometry that is related to unquenched flavors in the D$3$-D$3'$ intersection but we remained on a general level. It would be interesting to consider the probe D$3$-brane embedding in the background derived here for the special case of $S^5$ and study the spectra of meson states, in order to compare to the studies of the probe geometries \cite{Myers:2006qr, Arean:2006pk}. 

A natural extension of our work would be to consider the D$3$-D$3'$ system at finite temperature. Constructing solutions for supergravity backgrounds with smeared flavor D$3'$-branes corresponding to deconfining phases of the field theory does not seem to be a daunting task given the success in other flavor deformations of the ambient super Yang-Mills theories in (3+1) dimensions. The physics of quenched quarks in the D$3$-D$3'$ setup in the presence of a black hole was considered in \cite{Cottrell:2015kra}, where the flavor D$3'$-branes were treated as probes.

Additionally, we can consider a circle compactification along either the $x^2$ or $x^3$ directions and impose antiperiodic boundary conditions on the fermions. This will induce an explicit breaking of supersymmetry with a resulting cigar-like geometry. Studying such a background would teach us lessons on a quiver gauge theories possessing a mass gap, presumably also showing confining behavior. 

Another future avenue would be to compute the holographic entanglement entropy and compare the result to the computation in the geometry in the probe limit \cite{Jensen:2013lxa}, in order to better understand the effects of the smeared D-branes.

The addition of gauge fields is clearly a topic of great interest also from condensed matter perspective. The phase diagram will certainly be modified in the Veneziano limit and the dynamical effects from quarks running in loops will have many effects. For example, it is known that the holographic matter as described by brane intersections do not have a quasiparticle description at finite densities. In particular, for the (1+1)-dimensional fluid the Landau-Fermi description is insufficient. The holographic description of the defect D$3$-D$3'$ CFT in relation to the holographic description of quantum liquids was considered, e.g. in \cite{Hung:2009qk,Jokela:2015aha,Itsios:2016ffv}. One relevant aspect is to ask how is the propagation of the zero sound affected by the presence of backreaction.
Furthermore, it would be interesting to examine the description of $(1+1)$-dimensional p-wave superconductors based on the backgrounds derived here and understand whether they exhibit  qualitative differences compared to the probe-brane description \cite{Bu:2012qr}. 

We hope to report on some of these directions in future works.

\ack{
We are grateful to Oliver DeWolfe for discussions. N.~J. and J.~M.~P. have been supported in part by the Academy of Finland grant no. 1322307.  K.~C.~R.s work is supported in part by the U.S. Department of Energy (DOE), Office of Science, Office of High Energy Physics, under Award Number DE-SC0010005.  
}

\begin{appendices}

\newsec{On Sasaki-Einstein manifolds and their quivers}[app: special_SE5]

In the main body of the paper we worked with the most abstract parameterization of a five-dimensional Sasaki-Einstein space. Here, we briefly discuss four very interesting choices and the holographic field theories associated with those.

	\subsec{\texorpdfstring{The sphere - $S^5$}{The sphere - S5}}

We start by specifying the Sasaki-Einstein space to be the five-dimensional sphere.

\subsubsecstar{The geometry}

We will write the $S^5$ as a $U(1)$ bundle over the $\mathbb{CP}^2$ basis. In this parameterization 
\eqn{
ds^2_{S^5} = \left(d \tau + A \right)^2 +  ds^2_{\mathrm{KE}}		\,			,
}[eq S5_aux_01]
we have explicitly 
\eqn{
ds^2_{S^5} = \frac{1}{4} d \chi^2 + \frac{1}{4} \cos^2 \frac{\chi}{2} \left( (\omega^1)^2 + (\omega^2)^2 + \sin^2 \frac{\chi}{2} (\omega^3)^2 \right)  + \left(d\tau + \frac{1}{2} \cos^2 \left(\frac{\chi}{2} \right) \omega^3 \right)^2		\,			,
}[eq: S5_aux_02]
where in the above the $\omega^i$ are the left-invariant $SU(2)$ one-forms. They are given by: 
\eqna{
\omega^1		&=	\cos \psi d\theta + \sin \psi \sin \theta d \phi		\,		,	\\
\omega^2		&=	\sin \psi d\theta - \cos \psi \sin \theta d \phi		\,		,	\\
\omega^3		&=	d\psi + \cos \theta d\phi								\,		.
}[eq: S5_aux_03]
In order to have an explicit expression for the complex structure two-form, given by $J=\tfrac{1}{2}dA=E^{\hat{5}} \wedge E^{\hat{6}} + E^{\hat{7}} \wedge E^{\hat{8}}$, we need to specify the one-form basis components. They are given by
\eqna{
E^{\hat{5}}		&=			\frac{1}{2} \cos\left( \frac{\chi}{2} \right) \omega^1 																		\,		,			\\
E^{\hat{6}}		&=			\frac{1}{2} \cos\left( \frac{\chi}{2} \right) \omega^2																		\,		,			\\
E^{\hat{7}}		&=			\frac{1}{2} \cos\left( \frac{\chi}{2} \right) \sin\left( \frac{\chi}{2} \right) \omega^3									\,		,			\\
E^{\hat{8}}		&=			\frac{1}{2} d\chi																											\,		.			\\
E^{9}			&=			d\tau + \frac{1}{2} \cos^2 \left(\frac{\chi}{2} \right) \omega^3															\,		.
}[eq: one_form_S5]

\subsubsecstar{The field theory}

The dual field theory is the four-dimensional $\mathcal{N}=4$ SYM. In an $\mathcal{N}=1$ language we can write the matter content as a vector multiplet and three chiral superfields, $\Phi_a$ with $a=1,2,3$, that transform in the adjoint representation of the gauge group. The gauge group is $SU(N)$ and the interactions are encoded in the superpotential 
\eqn{
\mathrm{Tr} \left( \Phi_1 [\Phi_2, \Phi_3 ] \right)		\,			,
}[eq: s5_superpotential] 
which is cubic.

\subsec{\texorpdfstring{The conifold - $T^{1,1}$}{The conifold - T1,1}}

Here we wish to briefly describe the $T^{1,1}$ space and the associated holographic quiver theory.  

\subsubsecstar{The geometry}

The conifold is a Calabi-Yau threefold with a conical singularity. The metric element can be simply expressed as $ds^2_6 = dr^2 + r^2 ds^2_{T^{1,1}}$, with $T^{1,1}$ being the base of the Calabi-Yau cone. More specifically, it is the $\frac{SU(2) \times SU(2)}{U(1)}$ coset. Furthermore, it is a $U(1)$ bundle over an $S^2 \times S^2$ space. The five-dimensional $T^{1,1}$ is written as 
\eqn{
ds^2_{T^{1,1}} = \left(\frac{1}{3} d \tau + A \right)^2 +  ds^2_{\mathrm{KE}}		\,			,
}[eq T11_aux_1]
and we have, explicitly
 \eqn{
ds^2_{T^{1,1}} = \frac{1}{6} \sum^{2}_{i=1} \left(d\theta^2_i + \sin^2 \theta_i d \phi^2_i \right) + \frac{1}{9} \left(d\tau + \sum^2_{i=1} \cos \theta_i d\phi_i \right)^2 \,		.
}[eq: T11_aux2] 
In order to have an explicit expression for the complex structure two-form, given by $J=\tfrac{1}{2}dA=E^{\hat{5}} \wedge E^{\hat{6}} + E^{\hat{7}} \wedge E^{\hat{8}}$, we need to specify the one-form basis components. They are given by
\eqna{
E^{\hat{5}}		&=			\frac{1}{\sqrt{6}} \sin \theta_1 d\phi_1 																							\,		,			\\
E^{\hat{6}}		&=			\frac{1}{\sqrt{6}} d\theta_1																										\,		,			\\
E^{\hat{7}}		&=			\frac{1}{\sqrt{6}} \sin \theta_2 d\phi_2 																							\,		,			\\
E^{\hat{8}}		&=			\frac{1}{\sqrt{6}} d\theta_2																										\,		,			\\
E^{9}			&=			\frac{1}{3}\left( d\tau + \cos \theta_1 d\phi_1 + \cos \theta_2 d\phi_2 \right)														\,		.			
}[eq: one_form_T11]

\subsubsecstar{The field theory}

In order to delineate the basic features of the dual CFT, it is most convenient to consider the conifold as the locus in $\mathbb{C}^4$ described by 
\eqn{
z_1 z_2 = z_3 z_4			\,				,
}[eq: T11_aux_01]
which has a conical singularity at the origin, as it should. It is possible to find expressions that relate the holomorphic coordinates to the angles of $T^{1,1}$, which are given by 
\eqn{
\frac{z_1}{z_3} = \frac{z_4}{z_2} = e^{-i \phi_1} \tan \frac{\theta_1}{2}		\,			,		\qquad
\frac{z_1}{z_4} = \frac{z_3}{z_2} = e^{-i \phi_2} \tan \frac{\theta_2}{2}		\,			.
}[eq: T11_aux_02] 
Having described the relation between the angular parameterization of $T^{1,1}$ and the one in terms of holomorphic coordinates, we want to stress that there is another way of solving the conifold \cref{eq: T11_aux_01}. We can consider 
\eqn{
z_1 = A_1 B_1		\,		,		\qquad		z_1 = A_1 B_1		\,		,		\qquad
z_3 = A_1 B_2		\,		,		\qquad		z_4 = A_2 B_1		\,		.		 		
}[eq: T11_aux_03]
The dual superconformal field theory is an $\mathcal{N}=1$ gauge theory with the group $SU(N) \times SU(M)$ and includes four $\mathcal{N}=1$ chiral supermultiplets. These are identified with the coordinates $A_1, A_2, B_1, B_2$. The fields $A_1, A_2$ transform in the $(N,M)$ representation of the gauge group, while $B_1$ and $B_2$ are in the $(M,N)$ representation. All of them have an R-charge given by $\tfrac{1}{2}$ and the superpotential of the theory is: 
\eqn{
g \epsilon^{ij} \epsilon^{kl} \mathrm{Tr} \left(A_i B_k A_j B_l \right)		\,			,
}[eq: T11_aux_04]
with $g$ being a constant.

\subsec{\texorpdfstring{Cohomogeneity one - $Y^{p,q}$}{Cohomogeneity one - Yp,q}}

Here we wish to briefly describe the $Y^{p,q}$ manifolds and their holographic quiver descriptions.

\subsubsecstar{The geometry}

The five-dimensional $Y^{p,q}$ is written as 
\eqn{
ds^2_{Y^{p,q}} = \left(\frac{1}{3} d \tau + A \right)^2 +  ds^2_{\mathrm{KE}}		\,			,
}[eq Ypq_aux_1]
and explicitly we have 
\eqna{
ds^2_{Y^{p,q}} = \frac{1-cy}{6}(d\theta^2 + \sin^2 \theta d\phi^2) &+ \frac{1}{w(y)q(y)}dy^2 + \frac{1}{36} w(y)q(y) (d\beta + c \cos \theta d\phi)^2  \\
&+ \frac{1}{9} \left( d\tau - \cos \theta d\phi + y (d\beta + c \cos \theta d\phi)\right)^2				\,				,
}[eq: Ypq_aux2]
with the functions being given by: 
\eqn{
w(y) = \frac{2(a-y^2)}{1-cy} \,		,		\qquad		q(y) = \frac{a-3y^2+2cy^3}{a-y^2}		\,		.
}[eq: Ypq_fnctns]
In order to have an explicit expression for the complex structure two-form, given by $J=\tfrac{1}{2}dA=E^{\hat{5}} \wedge E^{\hat{6}} + E^{\hat{7}} \wedge E^{\hat{8}}$, we need to specify the one-form basis components. They are given by
\eqna{
E^{\hat{5}}			&=			\sqrt{\frac{1-cy}{6}} d\theta 																							\,		,			\\
E^{\hat{6}}			&=			\sqrt{\frac{1-cy}{6}} \sin \theta d\phi																					\,		,			\\
E^{\hat{7}}			&=			\frac{1}{\sqrt{w(y)q(y)}} dy																							\,		,			\\
E^{\hat{8}}			&=			\frac{\sqrt{w(y)q(y)}}{6} \left(d\beta + c \cos \theta d\phi \right)^2													\,		,			\\
E^{9}				&=			\frac{1}{3} \left( d\tau - \cos \theta d\phi + y \left( d\beta + c \cos \theta d\phi \right) \right)					\,		.
}[eq: one_form_Ypq]

\subsubsecstar{The field theory}

The superconformal quiver gauge theories that are dual to the $Y^{p,q}$ spaces were originally worked out in \cite{Benvenuti:2004dy, Martelli:2004wu}. The associated gauge theories possess an $SU(2N)^{2p}$ gauge group and there exists a global symmetry given by $SU(2)$. There exists, also, a global flavor symmetry given by $U(1)_F$, as well as a baryon symmetry $U(1)_B$. We can build the superpotential of the theory by considering the contributions from three different types of terms, which we write below
\eqn{
\epsilon_{\alpha \beta}~U^{\alpha}_{L}~V^{\beta}~Y 	\,		,	\qquad		\epsilon_{\alpha \beta}~U^{\alpha}_{R}~V^{\beta}~Y	\,		,	\qquad		\epsilon_{\alpha \beta}~Z~U^{\alpha}_{R}~Y~U^{\beta}_{L}		\,			,
}[eq: superpotential_Ypq]
where a trace over the color indices is implied. In total we have $p+q$ couplings in the expression for the superpotential. 

In \cref{table: Ypq_fields}, we present the charges for the bifundamental fields in the quivers. 
\begin{table}[htb]
\begin{center}
\begin{tabular}{|c|c|c|c|c|}
 \hline
 &&&&\\[-0.95em] 
 Field  		& R-charge 													& $U(1)_{F}$ 	& $U(1)_{B}$ 	 	& Number  		\\ 
 \hline
 $Y$ 			& $\frac{-4p^2+3q^2+2pq+(2p-q)\sqrt{4p^2-3q^2}}{3q^2}$ 		& $-1$ 			& $p-q$ 			& $p+q$  	\\
 \hline
 $Z$ 			& $\frac{-4p^2+3q^2-2pq+(2p+q)\sqrt{4p^2-3q^2}}{3q^2}$ 		& $1$ 			& $p+q$ 			& $p-q$ 		 	\\
 \hline
 $U$ 			& $\frac{4p^2-2p\sqrt{4p^2-3q^2}}{3q^2}$	 				& $0$ 			& $-p$ 				& $p$ 		 	\\
 \hline
 $V$ 			& $\frac{3q-2p+\sqrt{4p^2-3q^2}}{3q}$ 						& $1$ 			& $q$ 				& $q$ 		 	\\
 \hline
\end{tabular}
\caption{Fields and charges in the $Y^{p,q}$ quivers.}
\label{table: Ypq_fields}
\end{center}
\end{table}

\subsubsecstar{Special limits of the spaces}

From the cohomogeneity-one five-dimensional Sasaki-Einstein spaces, $Y^{p,q}$, we can obtain some interesting geometries by considering special values for the variables that parameterize the metric \cite{Gauntlett:2004yd}. We list them below: 
\begin{itemize}
		\item $S^5$: the round five-sphere can be obtained by setting $c=a=1$ above. The base of the sphere is $\mathbb{CP}^2$. 
		\item $T^{1,1}$: the conifold is the special case in which $a=3$ and $c=0$. For $c \neq 0$ we obtain the orbifold $T^{1,1}/\mathbb{Z}_2$. 
\end{itemize}

\subsec{\texorpdfstring{Cohomogeneity two - $L^{a,b,c}$}{Cohomogeneity two - La,b,c}}

Here we wish to briefly describe the $L^{a,b,c}$ manifolds and their holographic quiver descriptions.

\subsubsecstar{The geometry}

The five-dimensional $L^{a,b,c}$ is written as 
\eqn{
ds^2_{L^{a,b,c}} = \left( d \tau + A \right)^2 +  ds^2_{\mathrm{KE}},
}[eq Labc_aux_1]
with the four-dimensional K\"ahler-Einstein metric
\eqna{ 
ds^2_{\mathrm{KE}} = \frac{\rho^2 dx^2}{4 \Delta_{x}}  + \frac{\rho^2 d \theta^2}{\Delta_{\theta}} &+ \frac{\Delta_{x}}{\rho^2} \left( \frac{\sin^2 \theta}{\alpha} d \phi + \frac{\cos^2 \theta}{\beta} d \psi  \right)^2 \\
&+ \frac{\Delta_{\theta} \sin^2 \theta \cos^2 \theta}{\rho^2} \left[ \left(\frac{\alpha-x}{\alpha} \right) d \phi - \left(\frac{\beta-x}{\beta} \right) d \psi  \right]^2, 
}[eq: Labc_part1]
with the relevant quantities appearing above being given by
\eqna{
A		 		&= \left(\frac{\alpha-x}{\alpha} \right) \sin^2 \theta d \phi + \left(\frac{\beta-x}{\beta} \right) \cos^2 \theta d \psi, \\
\rho^2 			&= \Delta_{\theta} - x, \\
\Delta_{x} 		&= x(\alpha-x)(\beta-x) - \mu, \\ 
\Delta_{\theta} &= \alpha \cos^2 \theta + \beta \sin^2 \theta.
}[eq: Labc_part2]
As it turns out, the metrics depend only on two non-trivial parameters since we can set to any non-zero value any one of $\alpha, \beta, \mu$ just by performing a rescaling of the remaining ones. The principal orbits are described as $U(1) \times U(1) \times U(1)$ and hence they are toric. They are degenerate when we evaluate them on the roots of $\Delta_x =0$ and at the special points $\theta=0, \pi/2$. We additionally present the ranges of the coordinates: $0 \leq \theta \leq \pi/2$, $0 \leq \{\theta, \psi\} \leq 2\pi$ and the $x$-coordinate ranges from $x_1 \leq x \leq x_2$ with $x_{1,2}$ denoting the smallest roots of the equation $\Delta_x =0$. The coordinate $\tau$ has a period $0\leq \tau \leq \breve{\tau}$ and $\breve{\tau}$ is to be defined subsequently. The three roots of the equation $\Delta_x =0$ are related to the metric constants $\alpha, \beta$ and $\mu$ as shown below:
\eqna{ 
\mu = x_1 x_2 x_3	\,		, 		\qquad \alpha + \beta = x_1 + x_2 + x_3	\,		,	 \qquad \alpha \beta = x_1 x_2 + x_1 x_3 + x_2 x_3		\,			,
}[eq: manifolddef]
where in the above $x_3$ is the third root of the cubic equation we described previously.

We note that it  is possible to find relations for $x_1, x_2, \alpha, \beta$ in terms of the quantities $a, b, c, d$. They have been obtained previously, see \cite{Canoura:2006es}, however, we find it convenient and useful to repeat the analysis here. The normalized Killing vector fields are given by: 
\eqn{
\partial_{\phi}, \quad \partial_{\psi}, \qquad \ell_{i} = A_i \partial_{\phi} + B_i \partial_{\psi} + C_i \partial_{A} 
}
with $i$ being valued either $1$ or $2$ and also, 
\eqn{ 
A_i = \frac{\alpha C_i}{x_i - \alpha}	\,		,	\qquad B_i = \frac{\beta C_i}{x_i - \beta} \,		,		\qquad 	C_i = \frac{(\alpha - x_i)(\beta-x_i)}{2(\alpha+\beta)x_i-\alpha \beta - 3 x^2_i}
}[eq: defABC] 
Now, we are at a position to give the value $\breve{\tau}$ which is equal to
\eqn{
\breve{\tau} =2\pi \frac{k |C_1|}{b}, \quad k=\gcd(a,b),
}
and $d$ is defined to be:
\eqn{
d = a+b-c
}[eq: defd]
The constants $A_i, B_i, C_i$ are related to to the integers $a,b,c$ that characterize the $L^{a,b,c}$ geometry through the relations
\eqn{  
a A_1 + b A_2 + c 	=0 \,		,	\qquad	a B_1 + b B_2 +d 	=0 	\,		,	\qquad	a C_1 + b C_2 		=0		\,		.
}[eq: preliminaries1]
A consequence of \cref{eq: preliminaries1} is that the ratios of $A_1 C_2 - A_2 C_1$, $B_1 C_2-B_2 C_1$, $C_1$, and $C_2$ have to be rational. More specifically, it has been shown that 
\eqn{
\frac{c}{b} = \frac{A_1 C_2 - A_2 C_1}{C_1} 		\,		,		\qquad
\frac{d}{b} = \frac{B_1 C_2 - B_2 C_1}{C_1} 		\,		,		\qquad
\frac{b}{a} = - \frac{C_1}{C_2}
}[eq: defratios]
Using \cref{eq: manifolddef,eq: preliminaries1,eq: defratios} we can derive 
\eqna{
\frac{c}{b} &= \frac{x_1(x_3-x_1)}{x_2(x_3-x_2)} 											\,		,		\qquad
\frac{a}{c} = \frac{(\alpha - x_2)(x_3 - x_1)}{\alpha(\beta-x_1)} \\
\frac{c}{d} &= \frac{\alpha (\beta-x_1)(\beta-x_2)}{\beta (\alpha - x_1)(\alpha - x_2)}		\,		,		\qquad
\frac{c}{d} =\frac{\alpha(x_3-\alpha)}{\beta(x_3-\beta)}
}[eq: usefuleqn]
In order to have an explicit expression for the complex structure two-form, given by $J=\tfrac{1}{2}dA=E^{\hat{5}} \wedge E^{\hat{6}} + E^{\hat{7}} \wedge E^{\hat{8}}$, we need to specify the one-form basis components. They are given by
\eqna{
E^{\hat{5}}			&=			\frac{\rho}{\sqrt{\Delta_{\theta}}} d\theta 																							\,		,			\\
E^{\hat{6}}			&=			\frac{\sqrt{\Delta_{\theta}} \sin \theta \cos \theta}{\rho} \left(\frac{\alpha-x}{\alpha} d\phi - \frac{\beta-x}{\beta} d\psi \right)	\,		,			\\
E^{\hat{7}}			&=			\frac{\sqrt{\Delta_x}}{\rho} \left( \frac{\sin^2 \theta}{\alpha} d\phi + \frac{\cos^2 \theta}{\beta} d\psi \right)						\,		,			\\
E^{\hat{8}}			&=			\frac{\rho}{2 \sqrt{\Delta_x}} dx																										\,		.			\\
E^{9}				&=			d\tau + A																																\,		.
}[eq: one_form_Labc]

\subsubsecstar{The field theory}

The superconformal quiver gauge theories that are dual to the $L^{a,b,c}$ spaces were originally worked out in \cite{Benvenuti:2005ja, Franco:2005sm, Butti:2005sw}. the associated gauge theories possess $N_G=a+b$ gauge groups and a number of $N_F=a+3b$ fields in the bifundamental representation. There exists a global, flavor symmetry given by $U(1)^2_F$, which is manifested in the bulk as the subgroup of isometries that leave the Killing spinors invariant. There is an ambiguity in the flavor symmetries as they mix with the baryon symmetry, $U(1)_B$, of the theories. We can build the superpotential of the theory by considering the contributions from three different types of terms, which we write below 
\eqn{
\mathrm{Tr}~Y~U_1~U_2	\,		,	\qquad		\mathrm{Tr}~Y~U_2~V_1	\,		,	\qquad		\mathrm{Tr}~Y~U_1~Z~U_2		\,			,
}[eq: superpotential_Labc]
the first two of which are cubic and the last is a quartic. The associated R-charge is equal to two and they are uncharged under the baryonic and the global flavor symmetries. We can determine the number of terms of each kind by examining the multiplicities of the fields. The number of individual terms from \cref{eq: superpotential_Labc} that build the superpotential are respectively $2(b-c)$, $2(c-a)$, and $2a$. The total number of terms that appear is, therefore, given by $N_F-N_G$. In \cref{table: Labc_fields}, we present the charges for the bifundamental fields in the quivers. 
\begin{table}[htb]
\begin{center}
\begin{tabular}{|c|c|c|c|c|c|}
 \hline
 &&&&&\\[-0.95em] 
 Field  		& R-charge 							& $U(1)_{F_1}$ 	& $U(1)_{F_2}$ 	& $U(1)_{B}$ 	& Number  		\\ 
 \hline
 $Y$ 			& $\frac{2(x_3-x_1)}{3x_3}$ 		& $1$ 			& $0$ 			& $a$ 			& $b$  	\\ 
 \hline
 $Z$ 			& $\frac{2(x_3-x_2)}{3x_3}$ 		& $0$ 			& $k$ 			& $b$ 			& $a$ 		 	\\
 \hline
 $U_1$ 			& $\frac{2 \alpha}{3x_3}$	 		& $0$ 			& $l$ 			& $-c$ 			& $d$ 		 	\\
 \hline
 $U_2$ 			& $\frac{2 \beta}{3x_3}$ 			& $-1$ 			& $-(k+l)$ 		& $-d$ 			& $c$ 		 	\\
 \hline
 $V_1$ 			& $\frac{4x_3+2x_1-2\beta}{3x_3}$ 	& $0$ 			& $k+l$ 		& $b-c$ 		& $c-a$ 		 	\\
 \hline
 $V_2$ 			& $\frac{4x_3+2x_1-2\alpha}{3x_3}$ & $-1$ 			& $-l$ 			& $c-a$ 		& $b-c$ 		 	\\
 \hline
\end{tabular}
\caption{Fields and charges in the $L^{a,b,c}$ quivers.}
\label{table: Labc_fields}
\end{center}
\end{table}

\subsubsecstar{Special limits of the spaces}

From the most general five-dimensional Sasaki-Einstein manifolds, $L^{a,b,c}$, we can obtain some interesting geometries by considering special values for the variables that parameterize the metric \cite{Cvetic:2005ft}. We list them below: 
\begin{itemize}
		\item $S^5$: the round five-sphere can be obtained by setting $\mu=0$ in \cref{eq: Labc_part2} above. The base of the sphere is $\mathbb{CP}^2$. 
		\item $T^{1,1}$: the conifold is the special case in which $a=b=c=1$. In this limit, the two lowest roots, given by $x_1$ and $x_2$ are degenerate and the four-dimensional K\"ahler-Einstein geometry becomes $S^2 \times S^2$. 
		\item $Y^{p,q}$: the cohomogeneity one Sasaki-Einstein manifolds are obtained from this general case upon taking the limit $a+b=2c$, which in turn implies $\alpha=\beta$. At the level of explicit values for the parameters we have $p-q=a$, $p+q=b$, and $p=c$. 
\end{itemize}

\newsec{Metric components in the UV and in the IR}[app: metrics]
In this appendix we list the UV and IR values of the corrections to the line element associated to the first-order $\epsilon$ expansion for the two profiles studied. For the constant profile solutions, \cref{eq: flavor_prof_11_cnstnt}, we have: 
\eqna{
\mathcal{A}_{1,\mathrm{IR}} &=
\frac{4r(25c_1-q_0+20\log r)}{25\sqrt{R}} 
~\epsilon, \,  \ \ \ 
\mathcal{A}_{1,\mathrm{UV}}=
\frac{4r(25c_1-q_0+20\log r)}{25\sqrt{R}} 
~\epsilon  \,   ,   \\
\mathcal{A}_{2,\mathrm{IR}} &=\left(\frac{2c_1 r^2}{\sqrt{R}} 
+ \frac{4r(25c_1+24 q_0 + 20 q_0 \log r)}{25\sqrt{R}} \right) \epsilon,  \, \ 
\mathcal{A}_{2,\mathrm{UV}} =\left(\frac{2c_1 r^2}{\sqrt{R}} 
+ \frac{4r(25c_1+24 q_0 + 20 q_0 \log r)}{25\sqrt{R}} \right) \epsilon \,   ,   \\
\mathcal{A}_{3,\mathrm{IR}} &=
\frac{(-25c_1+q_0-20\log r)\sqrt{R}}{25r^3} 
~\epsilon, \ \ \,   
\mathcal{A}_{3,\mathrm{UV}}=
\frac{(-25c_1+q_0-20\log r)\sqrt{R}}{25r^3} 
~\epsilon \,   ,   \\
\mathcal{A}_{4,\mathrm{IR}} &= 
\frac{2}{25}q_0 r \sqrt{R} 
~\epsilon, \ \ \, 
\mathcal{A}_{4,\mathrm{UV}} = 
\frac{2}{25}q_0 r \sqrt{R} 
~\epsilon \,   ,   \\
\mathcal{A}_{5,\mathrm{IR}} &= 
- \frac{3q_0 \sqrt{R}}{25r} 
~\epsilon, \ \ \,   
\mathcal{A}_{5,\mathrm{UV}} = 
-\frac{3q_0 \sqrt{R}}{25r} 
~\epsilon  \,   .   \\
}[eq: fnctn_aux_01]

For the profile that has a non-trivial dependence on the holographic radial coordinate, \cref{eq: flavor_prof_11}, the analogous quantities are given by: 
\eqna{
\mathcal{A}_{1,\mathrm{IR}} &= \frac{2r^4}{3 \sqrt{R}}~\epsilon, \ \  \, 
\mathcal{A}_{1,\mathrm{UV}} = \frac{4r(20\log r -1)}{25 \sqrt{R}}~\epsilon \,   ,   \\
\mathcal{A}_{2,\mathrm{IR}} &= \frac{2(9 a_i + \sqrt{3} \pi)r^2}{9\sqrt{R}}~\epsilon \,, \ \     
\mathcal{A}_{2,\mathrm{UV}} = \left(\frac{16r(6+5\log r)}{25\sqrt{R}} - \frac{2(\sqrt{3} \pi - 3 a_i)}{3\sqrt{R}} \right) \epsilon \,   ,   \\
\mathcal{A}_{3,\mathrm{IR}} &= -\frac{\sqrt{R}}{6}~\epsilon \,, \ \     
\mathcal{A}_{3,\mathrm{UV}} = \frac{\sqrt{R}(1-20 \log r)}{25r^3}~\epsilon \,   ,   \\
\mathcal{A}_{4,\mathrm{IR}} &= \frac{1}{8} r^2 \sqrt{R}~\epsilon \,, \ \   
\mathcal{A}_{4,\mathrm{UV}} = \frac{2\sqrt{R}}{25r}~\epsilon \,   ,   \\
\mathcal{A}_{5,\mathrm{IR}} &= - \frac{2 \pi r^4 \sqrt{R}}{15 \sqrt{3}}~\epsilon \,, \ \    
\mathcal{A}_{5,\mathrm{UV}} = - \frac{3\sqrt{R}}{25r}~\epsilon \,   .   \\
}[eq: fnctn_aux_02]

\end{appendices}
\bibliography{d3d3}

\begin{thebibliography}{10}
\ifx\href\asklfhas\newcommand{\href}[2]{#2}\fi
\ifx\arxivref\asklfhas\newcommand{\arxivref}[2]{\href{http://arxiv.org/abs/#1}{#2}}\fi
\ifx\doiref\asklfhas\newcommand{\doiref}[2]{\href{http://dx.doi.org/#1}{#2}}\fi
\parskip 0pt
\normalsize

\bibitem{Maldacena:1997re}
J.~M. Maldacena,
\textit{``{The Large N limit of superconformal field theories and
  supergravity}''},
\doiref{10.1023/A:1026654312961}{Adv.~Theor.~Math.~Phys. \textbf{2}, 231
  (1998)\ignorespaces}\ignorespaces,
\normalsize{\texttt{\arxivref{hep-th/9711200}{hep-th/9711200}}}\ignorespaces
\bibitem{Witten:1998qj}
E.~Witten,
\textit{``{Anti-de Sitter space and holography}''},
\doiref{10.4310/ATMP.1998.v2.n2.a2}{Adv.~Theor.~Math.~Phys. \textbf{2}, 253
  (1998)\ignorespaces}\ignorespaces,
\normalsize{\texttt{\arxivref{hep-th/9802150}{hep-th/9802150}}}\ignorespaces
\bibitem{Gubser:1998bc}
S.~S. Gubser, I.~R. Klebanov \& A.~M. Polyakov,
\textit{``{Gauge theory correlators from noncritical string theory}''},
\doiref{10.1016/S0370-2693(98)00377-3}{Phys.~Lett.~B \textbf{428}, 105
  (1998)\ignorespaces}\ignorespaces,
\normalsize{\texttt{\arxivref{hep-th/9802109}{hep-th/9802109}}}\ignorespaces
\bibitem{Itzhaki:1998dd}
N.~Itzhaki, J.~M. Maldacena, J.~Sonnenschein \& S.~Yankielowicz,
\textit{``{Supergravity and the large N limit of theories with sixteen
  supercharges}''},
\doiref{10.1103/PhysRevD.58.046004}{Phys.~Rev.~D \textbf{58}, 046004
  (1998)\ignorespaces}\ignorespaces,
\normalsize{\texttt{\arxivref{hep-th/9802042}{hep-th/9802042}}}\ignorespaces
\bibitem{Fayyazuddin:1998fb}
A.~Fayyazuddin \& M.~Spalinski,
\textit{``{Large N superconformal gauge theories and supergravity
  orientifolds}''},
\doiref{10.1016/S0550-3213(98)00545-8}{Nucl.~Phys.~B \textbf{535}, 219
  (1998)\ignorespaces}\ignorespaces,
\normalsize{\texttt{\arxivref{hep-th/9805096}{hep-th/9805096}}}\ignorespaces
\bibitem{Aharony:1998xz}
O.~Aharony, A.~Fayyazuddin \& J.~M. Maldacena,
\textit{``{The Large N limit of N=2, N=1 field theories from three-branes in F
  theory}''},
\doiref{10.1088/1126-6708/1998/07/013}{JHEP \textbf{9807}, 013
  (1998)\ignorespaces}\ignorespaces,
\normalsize{\texttt{\arxivref{hep-th/9806159}{hep-th/9806159}}}\ignorespaces
\bibitem{Bertolini:2001qa}
M.~Bertolini, P.~Di~Vecchia, M.~Frau, A.~Lerda \& R.~Marotta,
\textit{``{N=2 gauge theories on systems of fractional D3/D7 branes}''},
\doiref{10.1016/S0550-3213(01)00568-5}{Nucl.~Phys.~B \textbf{621}, 157
  (2002)\ignorespaces}\ignorespaces,
\normalsize{\texttt{\arxivref{hep-th/0107057}{hep-th/0107057}}}\ignorespaces
\bibitem{Grana:2005sn}
M.~Grana, R.~Minasian, M.~Petrini \& A.~Tomasiello,
\textit{``{Generalized structures of N=1 vacua}''},
\doiref{10.1088/1126-6708/2005/11/020}{JHEP \textbf{0511}, 020
  (2005)\ignorespaces}\ignorespaces,
\normalsize{\texttt{\arxivref{hep-th/0505212}{hep-th/0505212}}}\ignorespaces
\bibitem{Karch:2002sh}
A.~Karch \& E.~Katz,
\textit{``{Adding flavor to AdS / CFT}''},
\doiref{10.1088/1126-6708/2002/06/043}{JHEP \textbf{0206}, 043
  (2002)\ignorespaces}\ignorespaces,
\normalsize{\texttt{\arxivref{hep-th/0205236}{hep-th/0205236}}}\ignorespaces
\bibitem{Karch:2000gx}
A.~Karch \& L.~Randall,
\textit{``{Open and closed string interpretation of SUSY CFT's on branes with
  boundaries}''},
\doiref{10.1088/1126-6708/2001/06/063}{JHEP \textbf{0106}, 063
  (2001)\ignorespaces}\ignorespaces,
\normalsize{\texttt{\arxivref{hep-th/0105132}{hep-th/0105132}}}\ignorespaces
\bibitem{Kruczenski:2003be}
M.~Kruczenski, D.~Mateos, R.~C. Myers \& D.~J. Winters,
\textit{``{Meson spectroscopy in AdS / CFT with flavor}''},
\doiref{10.1088/1126-6708/2003/07/049}{JHEP \textbf{0307}, 049
  (2003)\ignorespaces}\ignorespaces,
\normalsize{\texttt{\arxivref{hep-th/0304032}{hep-th/0304032}}}\ignorespaces
\bibitem{Myers:2006qr}
R.~C. Myers \& R.~M. Thomson,
\textit{``{Holographic mesons in various dimensions}''},
\doiref{10.1088/1126-6708/2006/09/066}{JHEP \textbf{0609}, 066
  (2006)\ignorespaces}\ignorespaces,
\normalsize{\texttt{\arxivref{hep-th/0605017}{hep-th/0605017}}}\ignorespaces
\bibitem{Arean:2006pk}
D.~Arean \& A.~V. Ramallo,
\textit{``{Open string modes at brane intersections}''},
\doiref{10.1088/1126-6708/2006/04/037}{JHEP \textbf{0604}, 037
  (2006)\ignorespaces}\ignorespaces,
\normalsize{\texttt{\arxivref{hep-th/0602174}{hep-th/0602174}}}\ignorespaces
\bibitem{Abt:2019tas}
R.~Abt, J.~Erdmenger, N.~Evans \& K.~S. Rigatos,
\textit{``{Light composite fermions from holography}''},
\doiref{10.1007/JHEP11(2019)160}{JHEP \textbf{1911}, 160
  (2019)\ignorespaces}\ignorespaces,
\normalsize{\texttt{\arxivref{1907.09489}{arXiv:1907.09489
  \![hep-th]}}}\ignorespaces
\bibitem{Nakas:2020hyo}
T.~Nakas \& K.~S. Rigatos,
\textit{``{Fermions and baryons as open-string states from brane junctions}''},
\doiref{10.1007/JHEP12(2020)157}{JHEP \textbf{2012}, 157
  (2020)\ignorespaces}\ignorespaces,
\normalsize{\texttt{\arxivref{2010.00025}{arXiv:2010.00025
  \![hep-th]}}}\ignorespaces
\bibitem{Kruczenski:2003uq}
M.~Kruczenski, D.~Mateos, R.~C. Myers \& D.~J. Winters,
\textit{``{Towards a holographic dual of large N(c) QCD}''},
\doiref{10.1088/1126-6708/2004/05/041}{JHEP \textbf{0405}, 041
  (2004)\ignorespaces}\ignorespaces,
\normalsize{\texttt{\arxivref{hep-th/0311270}{hep-th/0311270}}}\ignorespaces
\bibitem{Babington:2003vm}
J.~Babington, J.~Erdmenger, N.~J. Evans, Z.~Guralnik \& I.~Kirsch,
\textit{``{Chiral symmetry breaking and pions in nonsupersymmetric gauge /
  gravity duals}''},
\doiref{10.1103/PhysRevD.69.066007}{Phys.~Rev.~D \textbf{69}, 066007
  (2004)\ignorespaces}\ignorespaces,
\normalsize{\texttt{\arxivref{hep-th/0306018}{hep-th/0306018}}}\ignorespaces
\bibitem{DeWolfe:2001pq}
O.~DeWolfe, D.~Z. Freedman \& H.~Ooguri,
\textit{``{Holography and defect conformal field theories}''},
\doiref{10.1103/PhysRevD.66.025009}{Phys.~Rev.~D \textbf{66}, 025009
  (2002)\ignorespaces}\ignorespaces,
\normalsize{\texttt{\arxivref{hep-th/0111135}{hep-th/0111135}}}\ignorespaces
\bibitem{Constable:2002xt}
N.~R. Constable, J.~Erdmenger, Z.~Guralnik \& I.~Kirsch,
\textit{``{Intersecting D-3 branes and holography}''},
\doiref{10.1103/PhysRevD.68.106007}{Phys.~Rev.~D \textbf{68}, 106007
  (2003)\ignorespaces}\ignorespaces,
\normalsize{\texttt{\arxivref{hep-th/0211222}{hep-th/0211222}}}\ignorespaces
\bibitem{Erdmenger:2007cm}
J.~Erdmenger, N.~Evans, I.~Kirsch \& E.~Threlfall,
\textit{``{Mesons in Gauge/Gravity Duals - A Review}''},
\doiref{10.1140/epja/i2007-10540-1}{Eur.~Phys.~J.~A \textbf{35}, 81
  (2008)\ignorespaces}\ignorespaces,
\normalsize{\texttt{\arxivref{0711.4467}{arXiv:0711.4467
  \![hep-th]}}}\ignorespaces
\bibitem{Ramallo:2013bua}
A.~V. Ramallo,
\textit{``{Introduction to the AdS/CFT correspondence}''},
\doiref{10.1007/978-3-319-12238-0_10}{Springer~Proc.~Phys. \textbf{161}, 411
  (2015)\ignorespaces}\ignorespaces,
\normalsize{\texttt{\arxivref{1310.4319}{arXiv:1310.4319
  \![hep-th]}}}\ignorespaces
\bibitem{Seiberg:1994pq}
N.~Seiberg,
\textit{``{Electric - magnetic duality in supersymmetric nonAbelian gauge
  theories}''},
\doiref{10.1016/0550-3213(94)00023-8}{Nucl.~Phys.~B \textbf{435}, 129
  (1995)\ignorespaces}\ignorespaces,
\normalsize{\texttt{\arxivref{hep-th/9411149}{hep-th/9411149}}}\ignorespaces
\bibitem{Banks:1981nn}
T.~Banks \& A.~Zaks,
\textit{``{On the Phase Structure of Vector-Like Gauge Theories with Massless
  Fermions}''},
\doiref{10.1016/0550-3213(82)90035-9}{Nucl.~Phys.~B \textbf{196}, 189
  (1982)\ignorespaces}\ignorespaces
\bibitem{Bigazzi:2009bk}
F.~Bigazzi, A.~L. Cotrone, J.~Mas, A.~Paredes, A.~V. Ramallo \& J.~Tarrio,
\textit{``{D3-D7 Quark-Gluon Plasmas}''},
\doiref{10.1088/1126-6708/2009/11/117}{JHEP \textbf{0911}, 117
  (2009)\ignorespaces}\ignorespaces,
\normalsize{\texttt{\arxivref{0909.2865}{arXiv:0909.2865
  \![hep-th]}}}\ignorespaces
\bibitem{Conde:2016hbg}
E.~Conde, H.~Lin, J.~M. Penin, A.~V. Ramallo \& D.~Zoakos,
\textit{``{D3\textendash{}D5 theories with unquenched flavors}''},
\doiref{10.1016/j.nuclphysb.2016.11.016}{Nucl.~Phys.~B \textbf{914}, 599
  (2017)\ignorespaces}\ignorespaces,
\normalsize{\texttt{\arxivref{1607.04998}{arXiv:1607.04998
  \![hep-th]}}}\ignorespaces
\bibitem{Bigazzi:2005md}
F.~Bigazzi, R.~Casero, A.~L. Cotrone, E.~Kiritsis \& A.~Paredes,
\textit{``{Non-critical holography and four-dimensional CFT's with
  fundamentals}''},
\doiref{10.1088/1126-6708/2005/10/012}{JHEP \textbf{0510}, 012
  (2005)\ignorespaces}\ignorespaces,
\normalsize{\texttt{\arxivref{hep-th/0505140}{hep-th/0505140}}}\ignorespaces
\bibitem{Veneziano:1976wm}
G.~Veneziano,
\textit{``{Some Aspects of a Unified Approach to Gauge, Dual and Gribov
  Theories}''},
\doiref{10.1016/0550-3213(76)90412-0}{Nucl.~Phys.~B \textbf{117}, 519
  (1976)\ignorespaces}\ignorespaces
\bibitem{Casero:2006pt}
R.~Casero, C.~Nunez \& A.~Paredes,
\textit{``{Towards the string dual of N=1 SQCD-like theories}''},
\doiref{10.1103/PhysRevD.73.086005}{Phys.~Rev.~D \textbf{73}, 086005
  (2006)\ignorespaces}\ignorespaces,
\normalsize{\texttt{\arxivref{hep-th/0602027}{hep-th/0602027}}}\ignorespaces
\bibitem{Benini:2006hh}
F.~Benini, F.~Canoura, S.~Cremonesi, C.~Nunez \& A.~V. Ramallo,
\textit{``{Unquenched flavors in the Klebanov-Witten model}''},
\doiref{10.1088/1126-6708/2007/02/090}{JHEP \textbf{0702}, 090
  (2007)\ignorespaces}\ignorespaces,
\normalsize{\texttt{\arxivref{hep-th/0612118}{hep-th/0612118}}}\ignorespaces
\bibitem{Benini:2007gx}
F.~Benini, F.~Canoura, S.~Cremonesi, C.~Nunez \& A.~V. Ramallo,
\textit{``{Backreacting flavors in the Klebanov-Strassler background}''},
\doiref{10.1088/1126-6708/2007/09/109}{JHEP \textbf{0709}, 109
  (2007)\ignorespaces}\ignorespaces,
\normalsize{\texttt{\arxivref{0706.1238}{arXiv:0706.1238
  \![hep-th]}}}\ignorespaces
\bibitem{Conde:2011sw}
E.~Conde \& A.~V. Ramallo,
\textit{``{On the gravity dual of Chern-Simons-matter theories with unquenched
  flavor}''},
\doiref{10.1007/JHEP07(2011)099}{JHEP \textbf{1107}, 099
  (2011)\ignorespaces}\ignorespaces,
\normalsize{\texttt{\arxivref{1105.6045}{arXiv:1105.6045
  \![hep-th]}}}\ignorespaces
\bibitem{Conde:2011rg}
E.~Conde, J.~Gaillard \& A.~V. Ramallo,
\textit{``{On the holographic dual of $N=1$ SQCD with massive flavors}''},
\doiref{10.1007/JHEP10(2011)023}{JHEP \textbf{1110}, 023
  (2011)\ignorespaces}\ignorespaces,
\normalsize{\texttt{\arxivref{1107.3803}{arXiv:1107.3803
  \![hep-th]}}}\ignorespaces,
[Erratum: JHEP 08, 082 (2013)]\ignorespaces
\bibitem{Penin:2017lqt}
J.~M. Penin, A.~V. Ramallo \& D.~Zoakos,
\textit{``{Anisotropic D3-D5 black holes with unquenched flavors}''},
\doiref{10.1007/JHEP02(2018)139}{JHEP \textbf{1802}, 139
  (2018)\ignorespaces}\ignorespaces,
\normalsize{\texttt{\arxivref{1710.00548}{arXiv:1710.00548
  \![hep-th]}}}\ignorespaces
\bibitem{Jokela:2019tsb}
N.~Jokela, J.~M. Pen\'\i{}n, A.~V. Ramallo \& D.~Zoakos,
\textit{``{Gravity dual of a multilayer system}''},
\doiref{10.1007/JHEP03(2019)064}{JHEP \textbf{1903}, 064
  (2019)\ignorespaces}\ignorespaces,
\normalsize{\texttt{\arxivref{1901.02020}{arXiv:1901.02020
  \![hep-th]}}}\ignorespaces
\bibitem{Hoyos:2020zeg}
C.~Hoyos, N.~Jokela, J.~M. Pen\'\i{}n \& A.~V. Ramallo,
\textit{``{Holographic spontaneous anisotropy}''},
\doiref{10.1007/JHEP04(2020)062}{JHEP \textbf{2004}, 062
  (2020)\ignorespaces}\ignorespaces,
\normalsize{\texttt{\arxivref{2001.08218}{arXiv:2001.08218
  \![hep-th]}}}\ignorespaces
\bibitem{Arean:2008az}
D.~Arean, P.~Merlatti, C.~Nunez \& A.~V. Ramallo,
\textit{``{String duals of two-dimensional (4,4) supersymmetric gauge
  theories}''},
\doiref{10.1088/1126-6708/2008/12/054}{JHEP \textbf{0812}, 054
  (2008)\ignorespaces}\ignorespaces,
\normalsize{\texttt{\arxivref{0810.1053}{arXiv:0810.1053
  \![hep-th]}}}\ignorespaces
\bibitem{Hoyos:2021vhl}
C.~Hoyos, N.~Jokela, J.~M. Pen\'\i{}n, A.~V. Ramallo \& J.~Tarr\'\i{}o,
\textit{``{Risking your NEC}''},
\normalsize{\texttt{\arxivref{2104.11749}{arXiv:2104.11749
  \![hep-th]}}}\ignorespaces
\bibitem{Nunez:2010sf}
C.~Nunez, A.~Paredes \& A.~V. Ramallo,
\textit{``{Unquenched Flavor in the Gauge/Gravity Correspondence}''},
\doiref{10.1155/2010/196714}{Adv.~High~Energy~Phys. \textbf{2010}, 196714
  (2010)\ignorespaces}\ignorespaces,
\normalsize{\texttt{\arxivref{1002.1088}{arXiv:1002.1088
  \![hep-th]}}}\ignorespaces
\bibitem{Gubser:1998vd}
S.~S. Gubser,
\textit{``{Einstein manifolds and conformal field theories}''},
\doiref{10.1103/PhysRevD.59.025006}{Phys.~Rev.~D \textbf{59}, 025006
  (1999)\ignorespaces}\ignorespaces,
\normalsize{\texttt{\arxivref{hep-th/9807164}{hep-th/9807164}}}\ignorespaces
\bibitem{Sparks:2010sn}
J.~Sparks,
\textit{``{Sasaki-Einstein Manifolds}''},
\doiref{10.4310/SDG.2011.v16.n1.a6}{Surveys~Diff.~Geom. \textbf{16}, 265
  (2011)\ignorespaces}\ignorespaces,
\normalsize{\texttt{\arxivref{1004.2461}{arXiv:1004.2461
  \![math.DG]}}}\ignorespaces
\bibitem{Klebanov:1998hh}
I.~R. Klebanov \& E.~Witten,
\textit{``{Superconformal field theory on three-branes at a Calabi-Yau
  singularity}''},
\doiref{10.1016/S0550-3213(98)00654-3}{Nucl.~Phys.~B \textbf{536}, 199
  (1998)\ignorespaces}\ignorespaces,
\normalsize{\texttt{\arxivref{hep-th/9807080}{hep-th/9807080}}}\ignorespaces
\bibitem{Gauntlett:2004yd}
J.~P. Gauntlett, D.~Martelli, J.~Sparks \& D.~Waldram,
\textit{``{Sasaki-Einstein metrics on S**2 x S**3}''},
\doiref{10.4310/ATMP.2004.v8.n4.a3}{Adv.~Theor.~Math.~Phys. \textbf{8}, 711
  (2004)\ignorespaces}\ignorespaces,
\normalsize{\texttt{\arxivref{hep-th/0403002}{hep-th/0403002}}}\ignorespaces
\bibitem{Cvetic:2005ft}
M.~Cvetic, H.~Lu, D.~N. Page \& C.~Pope,
\textit{``{New Einstein-Sasaki spaces in five and higher dimensions}''},
\doiref{10.1103/PhysRevLett.95.071101}{Phys.~Rev.~Lett. \textbf{95}, 071101
  (2005)\ignorespaces}\ignorespaces,
\normalsize{\texttt{\arxivref{hep-th/0504225}{hep-th/0504225}}}\ignorespaces
\bibitem{Martelli:2005wy}
D.~Martelli \& J.~Sparks,
\textit{``{Toric Sasaki-Einstein metrics on S**2 x S**3}''},
\doiref{10.1016/j.physletb.2005.06.059}{Phys.~Lett.~B \textbf{621}, 208
  (2005)\ignorespaces}\ignorespaces,
\normalsize{\texttt{\arxivref{hep-th/0505027}{hep-th/0505027}}}\ignorespaces
\bibitem{Benvenuti:2004dy}
S.~Benvenuti, S.~Franco, A.~Hanany, D.~Martelli \& J.~Sparks,
\textit{``{An Infinite family of superconformal quiver gauge theories with
  Sasaki-Einstein duals}''},
\doiref{10.1088/1126-6708/2005/06/064}{JHEP \textbf{0506}, 064
  (2005)\ignorespaces}\ignorespaces,
\normalsize{\texttt{\arxivref{hep-th/0411264}{hep-th/0411264}}}\ignorespaces
\bibitem{Martelli:2004wu}
D.~Martelli \& J.~Sparks,
\textit{``{Toric geometry, Sasaki-Einstein manifolds and a new infinite class
  of AdS/CFT duals}''},
\doiref{10.1007/s00220-005-1425-3}{Commun.~Math.~Phys. \textbf{262}, 51
  (2006)\ignorespaces}\ignorespaces,
\normalsize{\texttt{\arxivref{hep-th/0411238}{hep-th/0411238}}}\ignorespaces
\bibitem{Benvenuti:2005ja}
S.~Benvenuti \& M.~Kruczenski,
\textit{``{From Sasaki-Einstein spaces to quivers via BPS geodesics:
  L**p,q|r}''},
\doiref{10.1088/1126-6708/2006/04/033}{JHEP \textbf{0604}, 033
  (2006)\ignorespaces}\ignorespaces,
\normalsize{\texttt{\arxivref{hep-th/0505206}{hep-th/0505206}}}\ignorespaces
\bibitem{Butti:2005sw}
A.~Butti, D.~Forcella \& A.~Zaffaroni,
\textit{``{The Dual superconformal theory for L**pqr manifolds}''},
\doiref{10.1088/1126-6708/2005/09/018}{JHEP \textbf{0509}, 018
  (2005)\ignorespaces}\ignorespaces,
\normalsize{\texttt{\arxivref{hep-th/0505220}{hep-th/0505220}}}\ignorespaces
\bibitem{Franco:2005sm}
S.~Franco, A.~Hanany, D.~Martelli, J.~Sparks, D.~Vegh \& B.~Wecht,
\textit{``{Gauge theories from toric geometry and brane tilings}''},
\doiref{10.1088/1126-6708/2006/01/128}{JHEP \textbf{0601}, 128
  (2006)\ignorespaces}\ignorespaces,
\normalsize{\texttt{\arxivref{hep-th/0505211}{hep-th/0505211}}}\ignorespaces
\bibitem{Giataganas:2009dr}
D.~Giataganas,
\textit{``{Semi-classical Strings in Sasaki-Einstein Manifolds}''},
\doiref{10.1088/1126-6708/2009/10/087}{JHEP \textbf{0910}, 087
  (2009)\ignorespaces}\ignorespaces,
\normalsize{\texttt{\arxivref{0904.3125}{arXiv:0904.3125
  \![hep-th]}}}\ignorespaces
\bibitem{Canoura:2005uz}
F.~Canoura, J.~D. Edelstein, L.~A. Pando~Zayas, A.~V. Ramallo \& D.~Vaman,
\textit{``{Supersymmetric branes on AdS(5) x Y**p,q and their field theory
  duals}''},
\doiref{10.1088/1126-6708/2006/03/101}{JHEP \textbf{0603}, 101
  (2006)\ignorespaces}\ignorespaces,
\normalsize{\texttt{\arxivref{hep-th/0512087}{hep-th/0512087}}}\ignorespaces
\bibitem{Canoura:2006es}
F.~Canoura, J.~D. Edelstein \& A.~V. Ramallo,
\textit{``{D-brane probes on L(a,b,c) Superconformal Field Theories}''},
\doiref{10.1088/1126-6708/2006/09/038}{JHEP \textbf{0609}, 038
  (2006)\ignorespaces}\ignorespaces,
\normalsize{\texttt{\arxivref{hep-th/0605260}{hep-th/0605260}}}\ignorespaces
\bibitem{Basu:2011di}
P.~Basu \& L.~A. Pando~Zayas,
\textit{``{Chaos rules out integrability of strings on AdS$_5 \times
  T^{1,1}$}''},
\doiref{10.1016/j.physletb.2011.04.063}{Phys.~Lett.~B \textbf{700}, 243
  (2011)\ignorespaces}\ignorespaces,
\normalsize{\texttt{\arxivref{1103.4107}{arXiv:1103.4107
  \![hep-th]}}}\ignorespaces
\bibitem{Basu:2011fw}
P.~Basu \& L.~A. Pando~Zayas,
\textit{``{Analytic Non-integrability in String Theory}''},
\doiref{10.1103/PhysRevD.84.046006}{Phys.~Rev.~D \textbf{84}, 046006
  (2011)\ignorespaces}\ignorespaces,
\normalsize{\texttt{\arxivref{1105.2540}{arXiv:1105.2540
  \![hep-th]}}}\ignorespaces
\bibitem{Rigatos:2020hlq}
K.~S. Rigatos,
\textit{``{Nonintegrability of $L^{a,b,c}$ quiver gauge theories}''},
\doiref{10.1103/PhysRevD.102.106022}{Phys.~Rev.~D \textbf{102}, 106022
  (2020)\ignorespaces}\ignorespaces,
\normalsize{\texttt{\arxivref{2009.11878}{arXiv:2009.11878
  \![hep-th]}}}\ignorespaces
\bibitem{Bea:2013jxa}
Y.~Bea, E.~Conde, N.~Jokela \& A.~V. Ramallo,
\textit{``{Unquenched massive flavors and flows in Chern-Simons matter
  theories}''},
\doiref{10.1007/JHEP12(2013)033}{JHEP \textbf{1312}, 033
  (2013)\ignorespaces}\ignorespaces,
\normalsize{\texttt{\arxivref{1309.4453}{arXiv:1309.4453
  \![hep-th]}}}\ignorespaces
\bibitem{Cottrell:2015kra}
W.~Cottrell, J.~Hanson, A.~Hashimoto, A.~Loveridge \& D.~Pettengill,
\textit{``{Intersecting D3-D3' -brane system at finite temperature}''},
\doiref{10.1103/PhysRevD.95.044022}{Phys.~Rev.~D \textbf{95}, 044022
  (2017)\ignorespaces}\ignorespaces,
\normalsize{\texttt{\arxivref{1509.04750}{arXiv:1509.04750
  \![hep-th]}}}\ignorespaces
\bibitem{Jensen:2013lxa}
K.~Jensen \& A.~O'Bannon,
\textit{``{Holography, Entanglement Entropy, and Conformal Field Theories with
  Boundaries or Defects}''},
\doiref{10.1103/PhysRevD.88.106006}{Phys.~Rev.~D \textbf{88}, 106006
  (2013)\ignorespaces}\ignorespaces,
\normalsize{\texttt{\arxivref{1309.4523}{arXiv:1309.4523
  \![hep-th]}}}\ignorespaces
\bibitem{Hung:2009qk}
L.-Y. Hung \& A.~Sinha,
\textit{``{Holographic quantum liquids in 1+1 dimensions}''},
\doiref{10.1007/JHEP01(2010)114}{JHEP \textbf{1001}, 114
  (2010)\ignorespaces}\ignorespaces,
\normalsize{\texttt{\arxivref{0909.3526}{arXiv:0909.3526
  \![hep-th]}}}\ignorespaces
\bibitem{Jokela:2015aha}
N.~Jokela \& A.~V. Ramallo,
\textit{``{Universal properties of cold holographic matter}''},
\doiref{10.1103/PhysRevD.92.026004}{Phys.~Rev.~D \textbf{92}, 026004
  (2015)\ignorespaces}\ignorespaces,
\normalsize{\texttt{\arxivref{1503.04327}{arXiv:1503.04327
  \![hep-th]}}}\ignorespaces
\bibitem{Itsios:2016ffv}
G.~Itsios, N.~Jokela \& A.~V. Ramallo,
\textit{``{Collective excitations of massive flavor branes}''},
\doiref{10.1016/j.nuclphysb.2016.06.008}{Nucl.~Phys.~B \textbf{909}, 677
  (2016)\ignorespaces}\ignorespaces,
\normalsize{\texttt{\arxivref{1602.06106}{arXiv:1602.06106
  \![hep-th]}}}\ignorespaces
\bibitem{Bu:2012qr}
Y.~Bu,
\textit{``{1+1-dimensional p-wave superconductors from intersecting
  D-branes}''},
\doiref{10.1103/PhysRevD.86.106005}{Phys.~Rev.~D \textbf{86}, 106005
  (2012)\ignorespaces}\ignorespaces,
\normalsize{\texttt{\arxivref{1205.1614}{arXiv:1205.1614
  \![hep-th]}}}\ignorespaces
\end{thebibliography}
\end{document}